\documentclass{aa}
\usepackage[utf8]{inputenc}
\usepackage{graphics}
\usepackage{txfonts}
\usepackage{amssymb}
\usepackage{lscape}
\usepackage{epsfig}
\usepackage{natbib}
\usepackage{longtable}
\usepackage{hyperref}
\usepackage[T1]{fontenc}
\usepackage[utf8]{inputenc}
\usepackage{multicol}
\usepackage[dvipsnames]{xcolor}

\title{The X-CLASS survey: A catalogue of 1646 X-ray-selected galaxy clusters up to z$\sim$1.5}
\author{E. Koulouridis\inst{1} \and N. Clerc\inst{2} \and T. Sadibekova\inst{3} \and M. Chira\inst{1,4} \and E. Drigga\inst{1,4} \and L. Faccioli\inst{3} \and J. P. Le Fèvre\inst{3} \and C. Garrel\inst{3} \and E. Gaynullina\inst{5} \and A. Gkini\inst{1,6} \and M. Kosiba\inst{7,8} \and F. Pacaud\inst{9} \and M. Pierre\inst{3} \and J. Ridl\inst{10} \and K. Tazhenova\inst{3} \and C. Adami\inst{11} \and B. Altieri\inst{12}  \and J.-C. Baguley\inst{13} \and R. Cabanac\inst{2} \and E. Cucchetti\inst{2}  \and A. Khalikova\inst{5} \and M. Lieu\inst{14}  \and J.-B. Melin\inst{15} \and M. Molham\inst{16} \and M. E. Ramos-Ceja\inst{10}   \and G. Soucail\inst{2} \and A. Takey\inst{16}   \and Ivan Valtchanov\inst{17}}
\institute{Institute for Astronomy \& Astrophysics, Space Applications \& Remote Sensing, National Observatory of Athens, GR-15236 Palaia Penteli, Greece
\and
IRAP, Université de Toulouse, CNRS, CNES, UT3, Toulouse, France
\and
AIM, CEA, CNRS, Universit\'e Paris-Saclay, Universit\'e Paris Diderot, Sorbonne Paris Cit\'e, F-91191 Gif-sur-Yvette, France
\and
Department of Physics, Aristotle University of Thessaloniki, Thessaloniki 54124, Greece
\and
Ulugh Beg Astronomical Institute of Uzbekistan Academy of Sciences, 33 Astronomicheskaya str., Tashkent, UZ-100052, Uzbekistan
\and
Department of Astrophysics, Astronomy \& Mechanics, Faculty of Physics, National and Kapodistrian University of Athens, Panepistimiopolis Zografou, 15784, Greece
\and
Department of Theoretical Physics and Astrophysics, Faculty of Science, Masaryk University, Kotl\'a\v rsk\'a 2, Brno, 611 37, Czech Republic
\and
Dipartimento di Fisica, Universit\`a degli Studi di Torino, via Pietro Giuria 1, I-10125 Torino, Italy
\and
Argelander-Institut f{\"{u}}r Astronomie, University of Bonn, Auf dem H{\"{u}}gel 71, D-53121 Bonn, Germany
\and
Max-Planck-Institut f{\"{u}}r extraterrestrische Physik, Giessenbachstra{\ss}e 1, D-85748 Garching, Germany
\and LAM, OAMP, Universit\'e Aix-Marseille, CNRS, P\^ole de l'\'Etoile, Site de Ch\~{a}teau Gombert, 38 rue Fr\'ed\'eric Joliot-Curie, 13388, Marseille 13 Cedex, France
\and
European Space Astronomy Centre, ESA, Villanueva de la Ca$\tilde{n}$ada, E-28691 Madrid, Spain
\and
School of Physics, HH Wills Physics Laboratory, Tyndall Avenue, Bristol, BS8 1TL, UK
\and
School of Physics \& Astronomy, University of Nottingham, University Park, Nottingham, NG7 2RD, UK
\and
IRFU, CEA, Universit{\'e} Paris-Saclay, F-91191 Gif-sur-Yvette, France
\and 
National Research Institute of Astronomy and Geophysics (NRIAG), 11421 Helwan, Cairo, Egypt
\and
Telespazio UK for ESA, European Space Astronomy Centre, Operations Department, 28691 Villanueva de la Ca\~nada, Spain
}
\date{\today}

\abstract{Cosmological probes based on galaxy clusters rely on cluster number counts and large-scale structure information. X-ray cluster surveys are well suited for this purpose because they are far less affected by projection effects than optical surveys, and cluster properties can be predicted with good accuracy.}{The XMM Cluster Archive Super Survey, X-CLASS, is a serendipitous search of X-ray-detected galaxy clusters in 4176 XMM-Newton archival observations until August 2015. All observations are clipped to exposure times of 10 and 20 ks to obtain uniformity, and they span $\sim$269 deg$^2$ across the high-Galactic latitude sky ($|b|> 20^o$). The main goal of the survey is the compilation of a well-selected cluster sample suitable for cosmological analyses.} {We describe the detection algorithm, the visual inspection, the verification process, and the redshift validation of the cluster sample, as well as the cluster selection function computed by simulations. We also present the various metadata that are released with the catalogue, along with two different count-rate measurements, an automatic one provided by the pipeline, and a more detailed and accurate interactive measurement. Furthermore, we provide the redshifts of 124 clusters obtained with a dedicated multi-object spectroscopic follow-up programme.}
{With this publication, we release the new X-CLASS catalogue of 1646 well-selected X-ray-detected clusters over a wide sky area, along with their selection function. The sample spans a wide redshift range, from the local Universe up to $z\sim$ 1.5, with 982 spectroscopically confirmed clusters, and over 70 clusters above $z=0.8$. The redshift distribution peaks at z$\sim0.1$, while if we remove the pointed observations it peaks at $z\sim0.3$. Because of its homogeneous selection and thorough verification, the cluster sample can be used for cosmological analyses, but also as a test-bed for the upcoming eROSITA observations and other current and future large-area cluster surveys. It is the first time that such a catalogue is made available to the community via an interactive database which gives access to a wealth of supplementary information, images, and data.}{}

 
\defcitealias{Clerc2012a}{CS12}          

\begin{document}

\maketitle
\section{Introduction}

\begin{figure*}
\centering
        \includegraphics[scale=0.63]{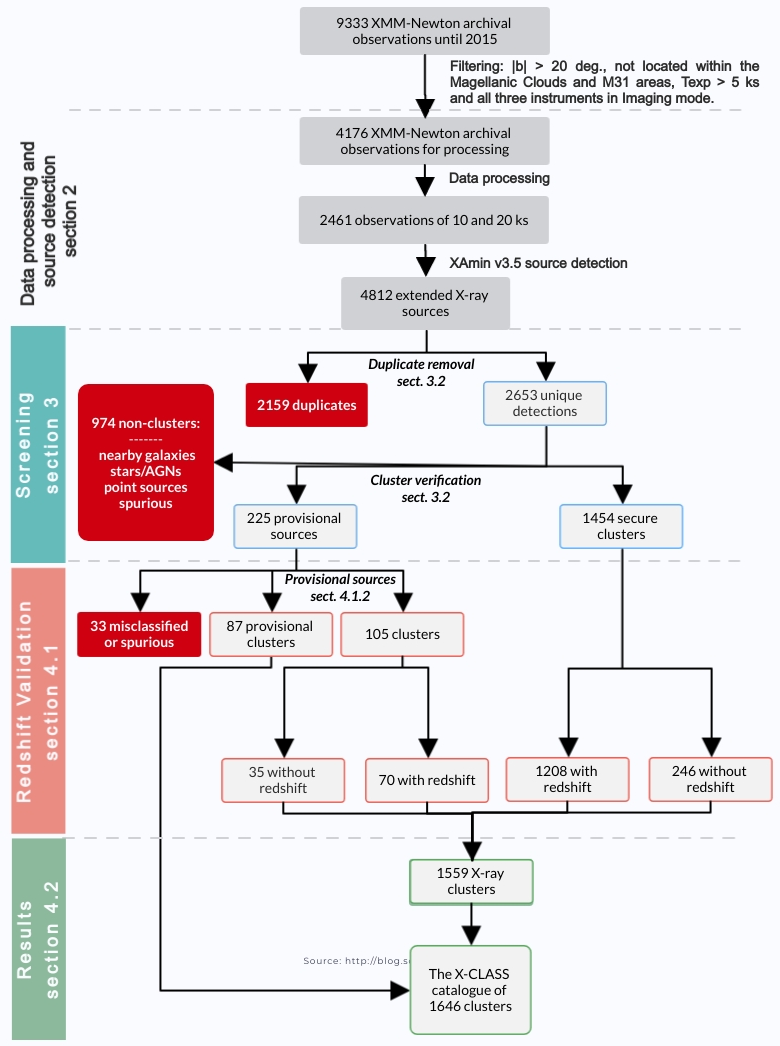}
        \caption{Flowchart of the overall procedure for the compilation of the new X-CLASS cluster catalogue. Red filled frames contain the sources that were discarded.}
        \label{fig:flow}
\end{figure*} 

\begin{figure*}
\centering
        \includegraphics[scale=0.85]{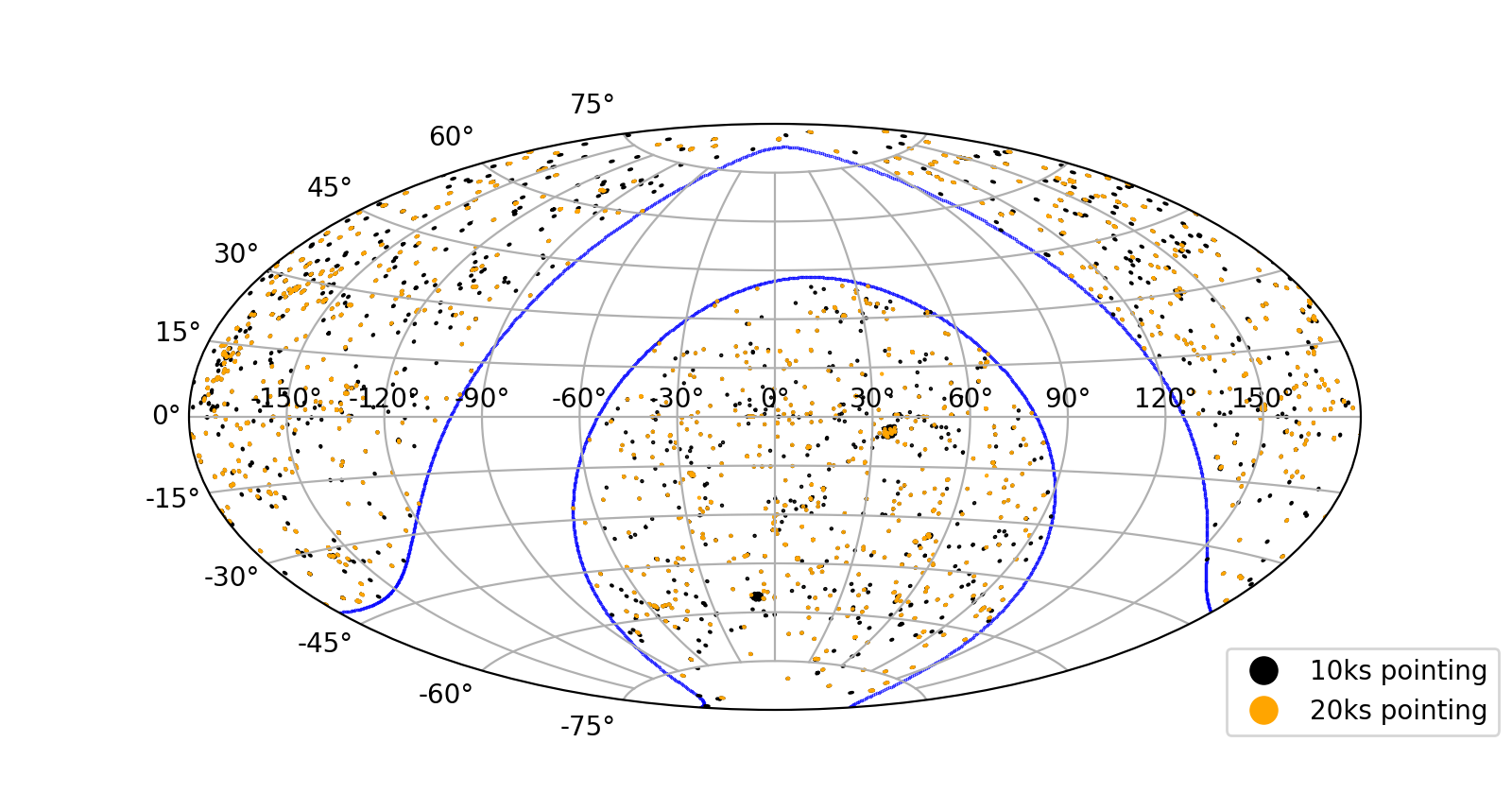}
        \includegraphics[scale=0.85]{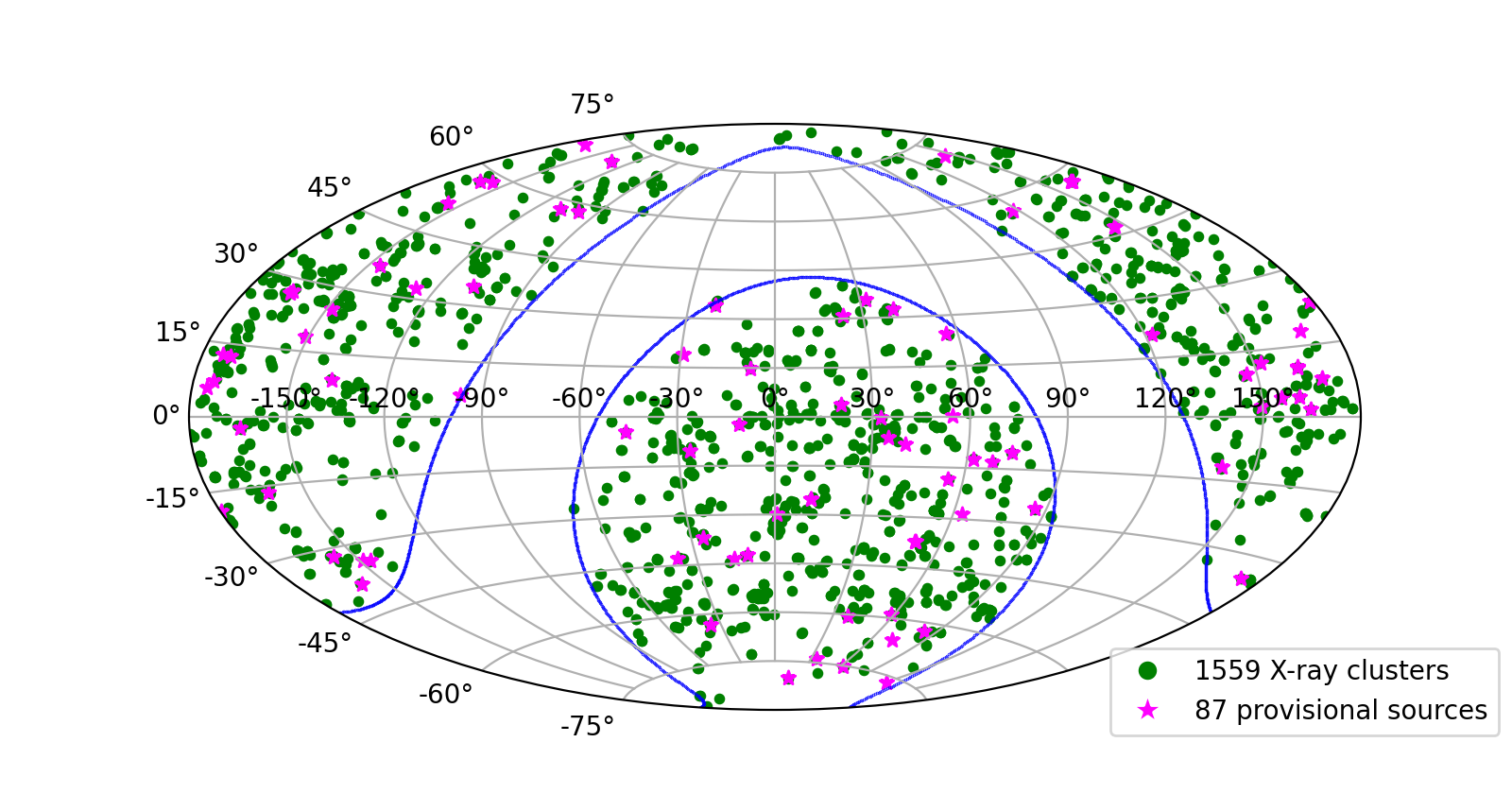}
        \caption{The X-CLASS survey covers a total area of 269.0 deg$^2$ by 4176 XMM-Newton archival observations. Top panel: there are 2461 "10 ks" pointings, of which 1309 also have a "20ks" version. This map in equatorial coordinates shows their location on the sky. The blue line marks the limit of $\pm$ 20 deg around the galactic plane. Bottom panel: the 1646 X-ray selected galaxy clusters of the X-CLASS catalogue.}
        \label{fig:cover}
\end{figure*} 

\begin{figure*}
\centering
        \includegraphics[scale=0.24]{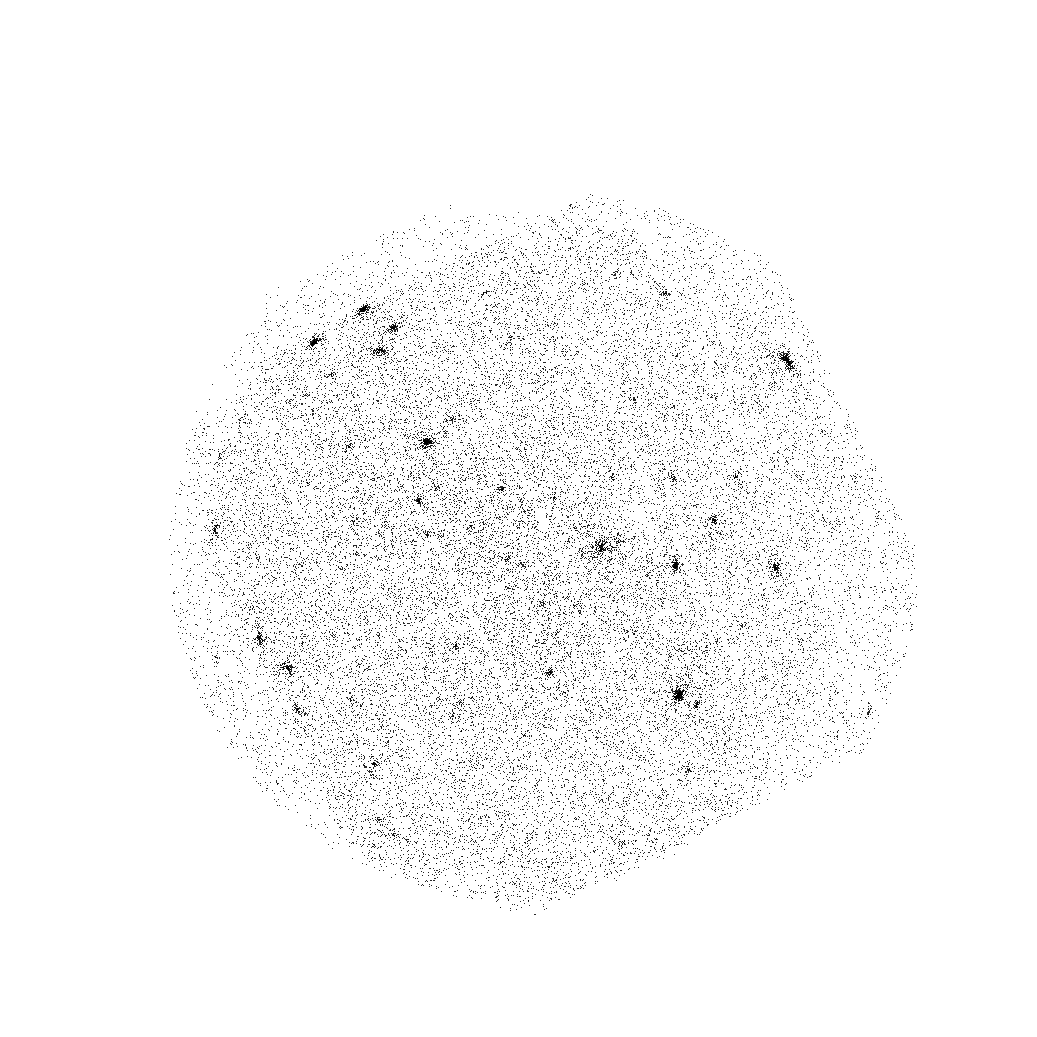}\includegraphics[scale=0.24]{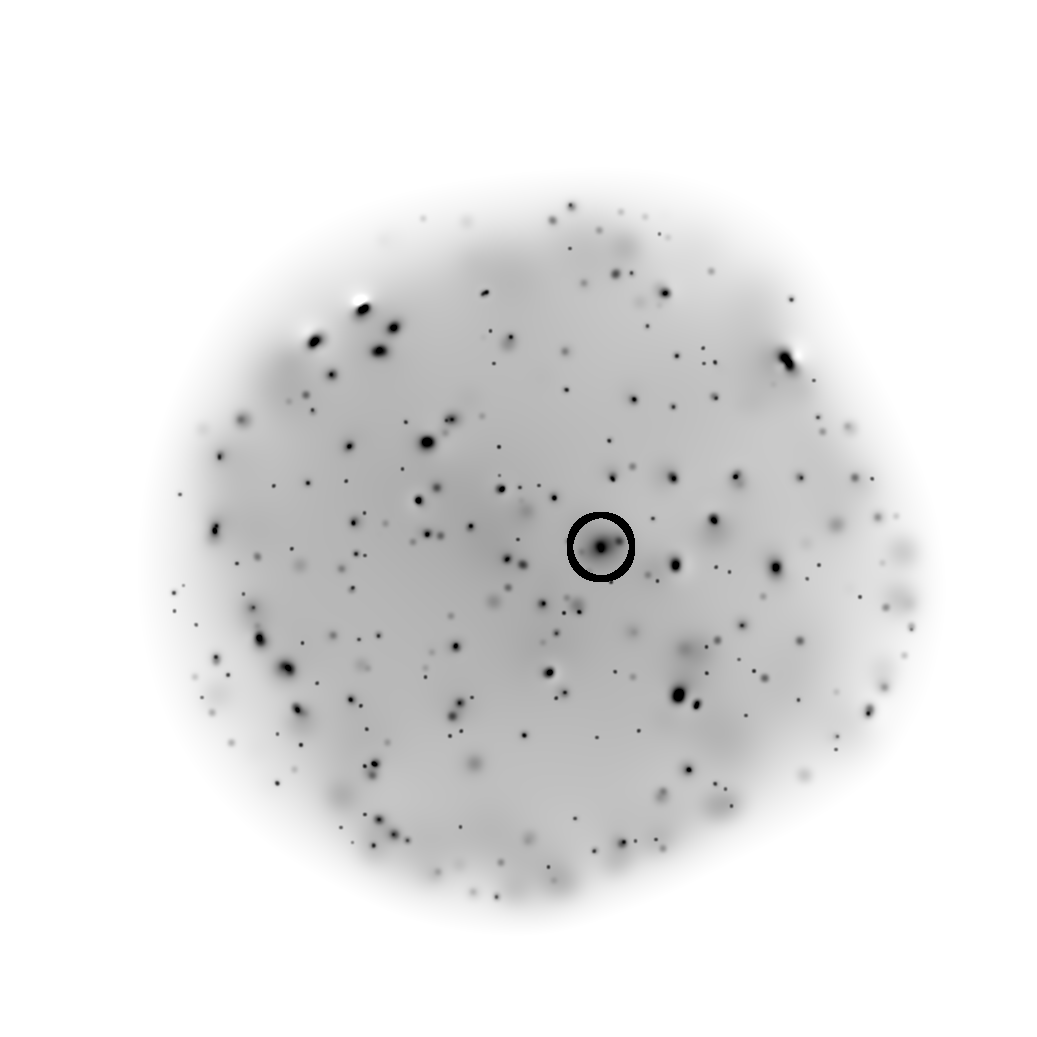}
        \caption{X-ray image in the [0.5 -- 2]\,keV X-ray band  of a 10 ks {\it XMM-Newton} observation (left panel, mosaic of all three XMM-newton detectors) and the corresponding wavelet filtered image (right panel). The images are not background-subtracted and their diameter is $26'$. The circle marks the position of a detected cluster candidate. Similar images are available for all X-CLASS clusters in the public database.}
        \label{fig:wavelet}
\end{figure*} 

Observational cosmology has been coming into increasing focus over the last two decades, propelled by wide-area surveys, such as for example the {\it Sloan Digital Sky Survey} \citep[SDSS;][]{Blanton2017} and the {\it Wide-field Infrared Survey Explorer} \citep[WISE;][]{Wright2010}, but also by modern observatories like {\it eROSITA} \citep{Merloni2012,Predehl2021} and {\it Euclid} \citep{Laureijs2011}. All these surveys gain additional value from supplemental multi-wavelength observations, with a growing synergy between space- and ground-based observatories. Wide-area surveys constitute an essential asset for large-scale structure studies because they can provide a unique handle on the abundance of massive objects, a crucial element for cosmology. Nevertheless, the collection, analysis, and treatment of the data, in order to create valuable user-friendly catalogues and archives for the scientific community, is increasingly challenging and time-consuming, despite the availability of modern tools and computational capabilities. Therefore, such endeavours are rightfully dubbed `legacy surveys'.

In this framework, cosmologists seek large samples of galaxy clusters, spanning a wide range of masses and redshifts, in order to use them as cosmological probes. These probes mainly rely on cluster number counts and large-scale structure information to capture the evolution of the halo mass function and the halo spatial distribution across cosmic times. Galaxy clusters are the most massive gravitationally bound structures in the Universe. They are mainly dominated by dark matter ($\sim$ 85\% of the total mass), while the hot X-ray-emitting intracluster medium (ICM) accounts for most of their baryonic mass \citep[e.g.][]{Plionis2008}. Therefore, galaxy clusters can be identified in various wavelengths \citep[e.g.][]{Vikhlinin2009, Mantz2010, Rozo2010, Sehgal2011, Benson2013, Planck2014}, and X-ray surveys in particular have proven very effective in detecting large numbers of them \citep[e.g.][]{Fassbender2011, Willis2013, Pierre2004, Pierre2016, Takey2016, Adami2018} including many at redshifts $z > 1$, with the most distant clusters found up to a redshift of $z\sim2$ \citep{Santos2011, Mantz2018}. 

The X-ray selection, although possibly biased towards baryon-rich and relaxed clusters \citep[e.g.][]{Andreon2016,O'Sullivan2017},  presents two main advantages: First, the cluster properties can be self-consistent and easily predicted by ab initio models, because the measurable X-ray parameters are closely related to the mass of the cluster \citep[e.g.][]{Frenk1990, vanHaarlem1997}. Second, cluster catalogues are hardly affected by projection effects \citep[e.g.][]{Ramos-Ceja2019}, namely the inclusion of spurious sources resulting from the projection of unrelated systems along the line of sight, because the centrally concentrated X-ray emission clearly indicates the presence of gas trapped in the potential well of a cluster \citep[e.g.][]{Frenk1990, Reblinsky1999}. 

Consequently, X-ray surveys have had a key role in the systematic search for galaxy clusters, initially with the historical HEAO-1 X-ray observatory \citep{Piccinotti1982} and then with the Einstein observatory Medium Sensitivity Survey \citep{Gioia1990, Henry1992}. Many other important surveys have been conducted over the course of the last three decades, starting with the ROSAT observatory, which provided the instrumentation for REFLEX-I, II \citep{Bohringer2001,Bohringer2014}, NORAS \citep{Bohringer2000}, MACS \citep{Ebeling2001}, ROSAT-NEP \citep{Henry2006} and CODEX \citep{Finoguenov2020, Kirkpatrick2021}, and then by XMM-Newton and Chandra observatories, which allowed COSMOS \citep{Scoville2007, Finoguenov2007}, XMM-LSS \citep{Pierre2007,Clerc2014}, XMM-BCS  \citep{Suhada2012},  XCS  \citep{Romer2001, Mehrtens2012},  2XMMi/SDSS \citep{Takey2011, Takey2013, Takey2014}, X-CLASS \citep[][hereafter CS12]{Clerc2012a}, XDCP \citep{Fassbender2011}, and XMM-XXL \citep{Pierre2016,Adami2018}.

In this context, here we present the new X-CLASS catalogue of 1646 X-ray selected clusters followed by a public release. The catalogue is based on all 9333 XMM archival observations publicly available until August 2015. All observations were filtered following identical criteria to those in \citetalias{Clerc2012a}: galactic latitudes above $20 \deg$ and not located within the Magellanic Clouds and M31 areas,  on sky exposure time larger than 5\,ks in each of the three European  Photon  Imaging  Camera (EPIC) detectors, and all three instruments in Imaging mode. This led to the selection of 4176 XMM pointings that went through an identical processing. In Sect. 2 we present the data processing, the resulting sample of galaxy cluster candidates, and their selection function. In Sect. 3 and 4 we describe the various steps of the screening of the candidates and further analysis of the sample. Finally, in Sect. 5 we describe the public database and in Sect. 6 we discuss the results and compare with other cluster catalogues. The full procedure is also illustrated in Fig.~\ref{fig:flow} in a flowchart. Throughout this paper, we use $H_0=70$ km s$^{-1}$ Mpc$^{-1}$, $\Omega_m=0.3$, and $\Omega_{\Lambda}=0.7$.

\section{Data processing and source detection}

The data processing follows the lines of \citetalias{Clerc2012a} with improvements and modifications. We recall here the main steps involved and highlight the main changes. Basic parameters used during the various data processing steps can be found in Table \ref{tab:processing}.

The XMM observations were reduced with the latest calibration files available (August 2015) and light curves were created in the high-energy band to monitor the flaring time-intervals. An automated algorithm identified excess variance in the count-rate time histogram in order to determine acceptable count-rate levels which were later included in the GTIs (good time intervals). More specifically, event  lists  were filtered from proton and solar flares by creating the high-energy event light curves in the [12–14] keV band for MOS and the [10–12] keV band for PN, and flagging out periods of high event rates (rates greater than $3\sigma$ above the mean observation count rate). Event histograms and the corresponding light curves and cut limits are stored in the online database and are available for all pointings. While the above automated procedure is efficient in removing short periods of high flares, it may fail in observations with a high mean particle background. Therefore, although no systematic human verification is performed at this point, parameters and figures were stored in a database and the overall quality of each observation was subsequently inspected by eye during the screening procedure described in sect. 3.3.
    
The 4176 observations were homogenised by selecting intervals not contaminated by flares, such that their exposure time in each of the EPIC detectors amounts exactly to 10\,ks or 20\,ks. This ensures uniformity, enabling the calculation of selection functions. In what follows we refer to XMM observations associated to one of the above two versions as `pointings'. Therefore, each original XMM observation may deliver zero, one (10 ks), or two (10 ks and 20 ks) X-CLASS pointings. This resulted in a total of 2461 observations with a 10 ks exposure, of which 1309  also had a 20 ks version. The distribution of these pointings on the sky is shown in Fig.~\ref{fig:cover}.  A circular area of radius $13 \arcmin$ around each pointing centre defines the geometric area of a pointing on the sky; sources are detected within this off-axis range only, thereby avoiding strongly vignetted areas on the detectors. The total survey geometrical area is computed by means of two-dimensional Monte-Carlo integration, accounting for the overlaps between pointings, and amounts to $269~\deg^2$. We note that pointed observations of clusters have been included in our observation list. Finally, they were processed twice with XAmin v3.5 pipeline \citep{faccioli2018} for both exposure times.
    
Source detection is a three-step process. During the first pass, images in the soft X-ray band, $[0.5-2]$\,keV, are created and filtered with a wavelet algorithm, as described in \citet{starck1998}. According to that paper, this is the best filtering method for X-ray images that contain few photons and Poisson noise. Most importantly, it was demonstrated that this method is very effective in detecting low-flux extended X-ray sources, which is crucial for our short-exposure time survey. An example of such a filtered observation is presented in Fig.~\ref{fig:wavelet}

\begin{table}
    \centering
    \caption{Data processing parameters}
    \begin{tabular}{lc}
         Parameter & value  \\
         \hline
         &\\
         {\bf Event selection:}&\\
         &\\
        MOS event flag selection & \#XMMEA\_EM \\
        PN event flag selection & (FLAG \& 0x2fb002c)==0\\
        MOS patterns & [0:12]\\
        PN patterns & [0:4]\\
        &\\
        {\bf Image}:&\\
        &\\
        Type &Sky\\
        Configuration &Co-addition of EPIC detectors\\
        Pixel size &2.5 arcsec\\
        Filtering &Wavelet transform (8 scales)\\
        &\\
        {\fontfamily{cmr}\selectfont {\bf SExtractor}}:&\\
        &\\
        Background cell side  &64 pixel\\
        Background median filtering  &4 cells\\
        Detection threshold  &$6\sigma$\\
        Detection minimum area  &12 pixel\\ 
        Deblending sub thresholds  &64\\
        Deblend min. contrast & 0.003\\
        &\\
        \hline
    \end{tabular}
    \label{tab:processing}
\end{table}

During a second step, {\fontfamily{cmr}\selectfont SExtractor} \citep{Bertin96} was used for source detection. It provided a centroid estimate, the extent of the X-ray emission, and a rough measurement of brightness. The background level is iteratively estimated in image cells by 3$\sigma$ clipping and a full-resolution background map is constructed by bicubic-spline interpolation. Finally, these parameters served as input for a maximum likelihood fitting routine that applied several source models on the photon image in the soft X-ray band: in particular, a precise point-spread function (PSF) model and an extended $\beta$-model \citep{cavaliere1978} described by \[S_X(r)\varpropto \left[1+\left(\frac{r}{EXT}\right)^2\right]^{-3\beta+1/2},\] where $EXT$ is the core radius in arc-seconds and $\beta=2/3$. For either a point-like or an extended source, the MEDIUM PSF model from the {\it XMM} calibration data is used. The {\fontfamily{cmr}\selectfont SExtractor} pixel segmentation mask was used to flag out pixels belonging to neighbouring sources included in the box. In contrast to previous XAmin versions, XAmin v3.5 fixed the position of the extended source fit at the value found by {\fontfamily{cmr}\selectfont SExtractor}. This selection yields 4812 extended X-ray sources.

Only sources classified as "C1" in a pointing were used to build the catalogue. These are characterised by a value of their extended-detection likelihood greater than 32, an extent likelihood greater than 33, and an extent (best-fit core-radius) larger than $5''$ \citep[for more details on these quantities see][]{faccioli2018}. Simulations of XMM `empty' cosmological fields demonstrated that such thresholds ensure pure samples of extended sources with a controllable selection function \citep{pacaud2006}. However, XMM-Newton is generally not pointing towards empty fields. The diversity of sources and instrumental artefacts encountered in the archive makes the X-CLASS C1 sample more prone to contamination by non-cluster sources and this is the purpose of the visual inspection process to eliminate those spurious sources (see Sect.~\ref{sect:visu_inspection}). At this stage, a given C1 detection may appear several times in the source list. This is either due to overlaps between XMM observations, because it appears in both exposure-time versions of the same pointing, or because it has been separated in multiple components, usually by sharp (small-scale) exposure variations due to one (or more) CCD gaps that caused the detection pipeline to segment the detection of extended sources into multiple areas. In the following sections, we describe the cluster selection function and the decontaminating procedure for the removal of duplicate and false detections. We note that clusters detected on observations with missing detectors were discarded from our list, but can be found in Table \ref{tab:discarded}.

Owing to the uniform exposure time, simulations allow us to accurately determine the selection function of the X-CLASS cluster sample. However, a large number of extended X-ray detections can be due to various other sources, such as for example nearby galaxies, stars, active galactic nuclei (AGNs), double point-sources, and so on. Most of these sources are not modelled in the simulations and should be discarded before reaching a clean cluster sample. We describe this interactive procedure in the following section.

\section{Screening of the X-ray cluster candidates}

\subsection{Previous versions of the catalogue}

Our previous analysis of the XMM archive provided a catalogue of 845 sources identified as galaxy clusters among 1514 C1 unique detections on which we performed visual screening. This first catalogue was publicly released in \citetalias{Clerc2012a}. \citet{Sadibekova2014} cross-matched the redMaPPer optical cluster catalogue \citep{Rykoff2014} in the northern hemisphere (SDSS photometry) and a subset of the X-CLASS C1 sources. These sources were included as targets of the SPIDERS ({\it SPectroscopic IDentifcation of ERosita Sources}) follow-up programme \citep{Clerc2016} in SDSS-IV \citep{Blanton2017} and are further described in Sect.~\ref{sect:spiders} . In addition, a large follow-up programme with the GROND instrument on  the  MPG/2.2m telescope at La Silla (Greiner et al. 2008) measured 232 photometric redshifts of X-CLASS clusters using 4-band images and is described in \citet{Ridl2017}. All 845 clusters of the first data release are included in the current new X-CLASS catalogue.

\subsection{Duplicate removal}

The initial list of C1 pipeline-detected sources requires a first-pass inspection in order to remove duplicate entries of single sources. An iterative procedure is performed, coupling automatic and manual associations of multiple detections. Starting from the list of 4812 C1 detections, we grouped sources with a fixed $10\arcsec$ distance criterion, reminiscent of the XMM-Newton PSF extent. We then ensured that the association flags are correctly set, reflecting the nature of the duplicate (e.g. overlapping pointing, secondary exposure, etc.). Only one entry is selected as the main detection according to the following rules: it needs to be detected on the higher exposure pointing (20\,ks if present) and in case of conflict, to have the higher value of extent likelihood. This procedure was repeated within $20''$, $30''$, $40''$, and $60''$ correlation radii. Correlations within the larger two radii were supervised and visually validated in order to prevent false associations. During this step, 2159 sources were identified as duplicate detections, leaving 2653 for further visual inspection.

\begin{figure*}
\centering
        \includegraphics[scale=0.42]{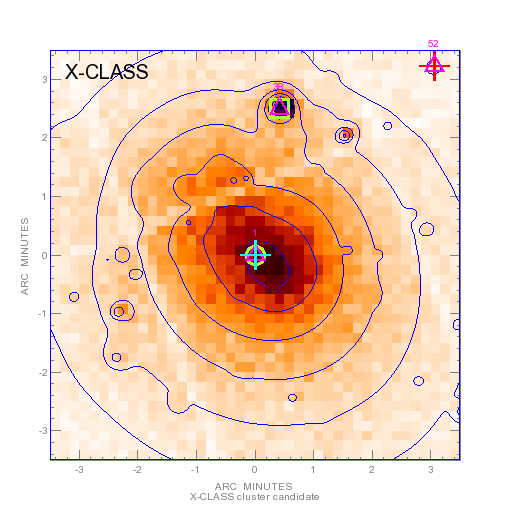}\vspace{0.2in}\hspace{0.25in}\includegraphics[scale=0.33]{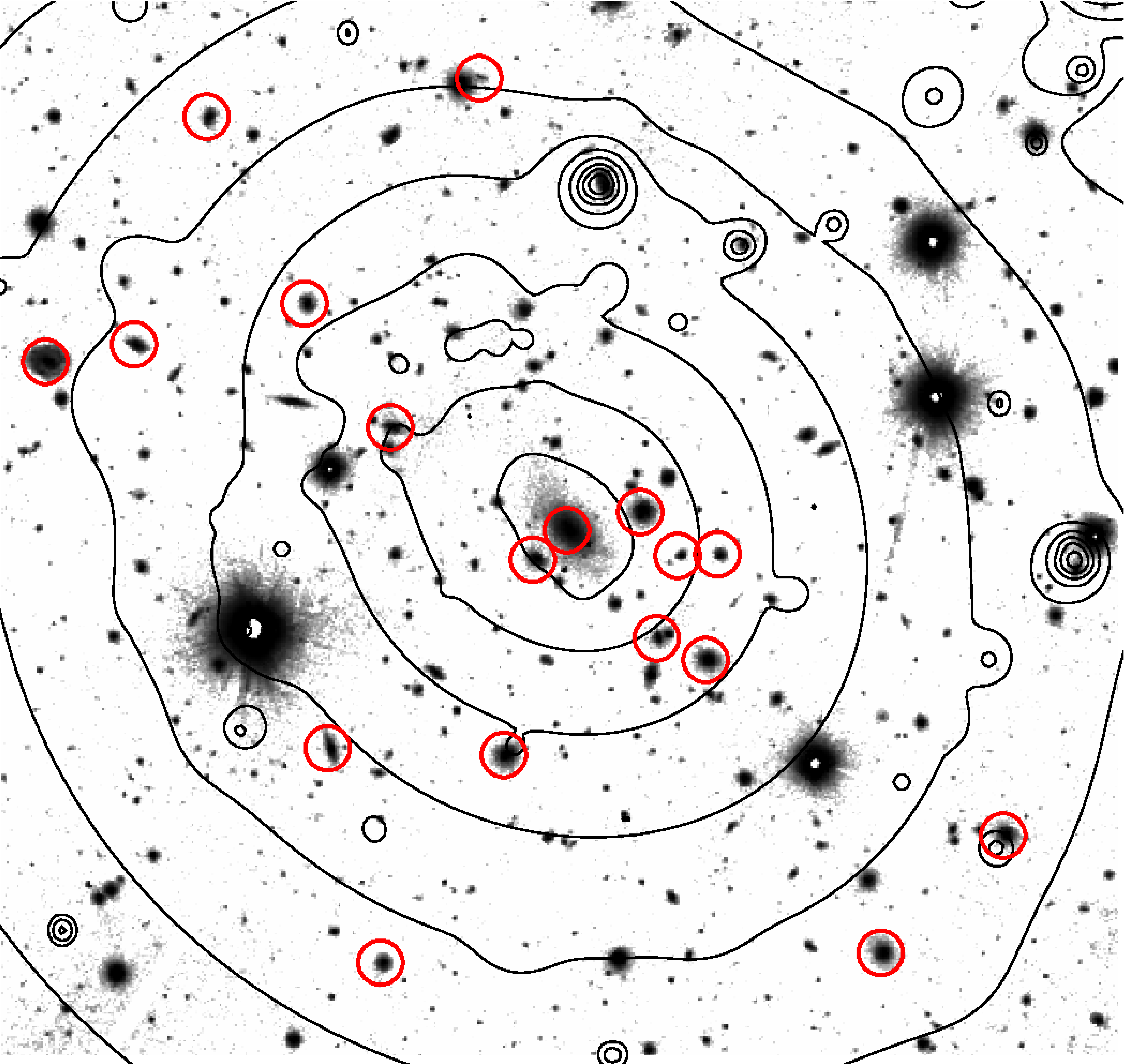}
        \includegraphics[scale=0.3]{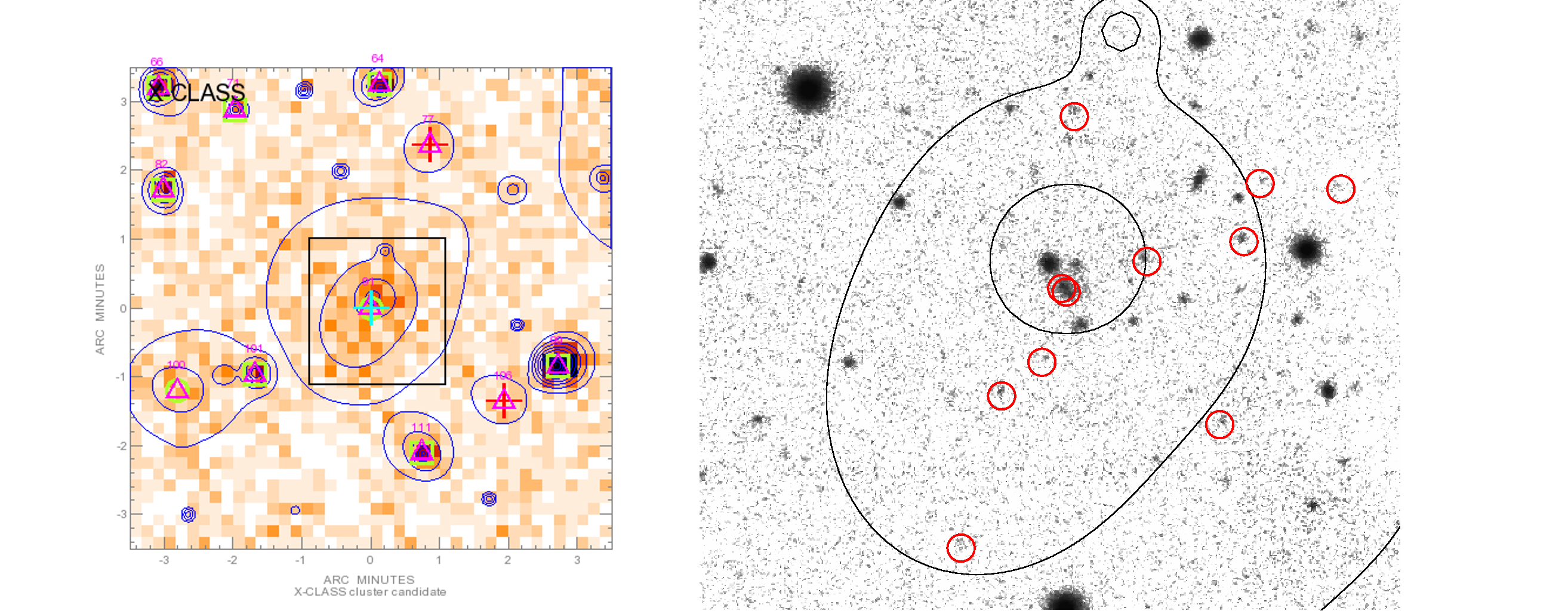}
        \caption{Examples of X-ray-selected galaxy clusters in the X-CLASS survey. Top panels: Cluster Xclass0561 (ABELL 2050) at z=0.119 as confirmed by 19 member galaxies. Bottom panels: Cluster Xclass0219 at z=0.791 as confirmed by 11 member galaxies. Left panels: X-ray images and contours. Green circles (squares) mark detections of extended (point-like) sources as classified by the XAmin pipeline. Straight lines that cross the image are CCD gaps of the XMM-Newton detector. Right panels: i-band optical images from PanSTARRS over-plotted with X-ray contours. Red circles mark the member galaxies with available spectroscopic redshift. In the case of Xclass0561, both X-ray and optical images cover the same sky area, while in the case of Xclass0219 the optical image corresponds to the central region of the X-ray image marked with the black square.}
        \label{fig:3380}
\end{figure*}

\subsection{Cluster verification\label{sect:visu_inspection}}

Once the duplication check had been applied, all C1 candidates were interactively screened by
expert researchers to identify non-clusters, spurious detections, or duplicates missed by the
algorithm. To this end, we used overlays of X-ray contours on the Digitized Sky Survey (DSS) images using the dedicated database tool. The purpose of this procedure is twofold: (1) remove nearby galaxies, saturated point-sources, X-ray artefacts, and possible unresolved double-sources that also appear as extended sources, and (2) provide an approximate distance indicator depending on the existence of a conspicuous optical counterpart to the X-ray emission, namely: NEARBY ($z<0.3-0.4$) and DISTANT ($z>0.3-0.4$), where $z \sim 0.3-0.4$ corresponds to the POSS-II plate limit for typical cluster red-sequence galaxies. Other classifications also included `nearby galaxy' and `fossil group'. The procedure involved several researchers from the X-CLASS collaboration specialised in cluster science and astronomical observations. Each source was inspected by two or more of the researchers who independently decided on its classification. The final classification, based on these decisions, was then assigned to the sources by two expert moderators, who did not participate in the classification. If the classifications provided by the researchers were discrepant, the decision was arbitrated by the moderators. During this step, 974 detections were discarded. All available redshifts were retrieved for the remaining 1679 cluster candidates using the NED extragalactic database and dedicated follow-up observations.

Out of the 1679 sources, 225 were flagged as `provisional', signifying that the nature of the detected source was dubious, and further investigation was needed. The nature of these sources was thoroughly scrutinised at a later time using deeper optical surveys and more current spectroscopic and photometric data not available during the initial catalogue compilation. Their screening is presented in more detail in Sect.~\ref{sect:dubius}.

\begin{figure*}
        \includegraphics[scale=0.7]{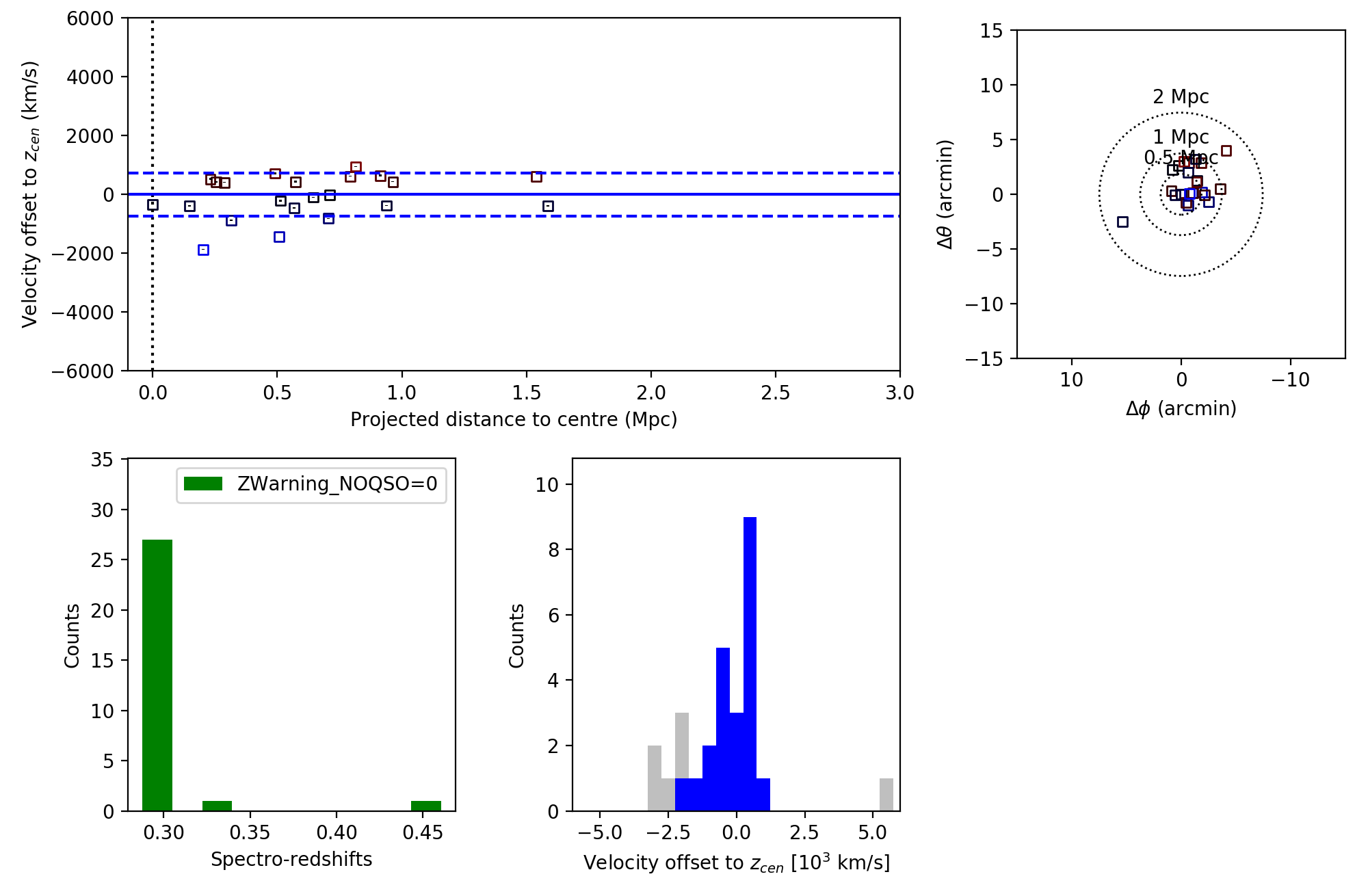}
        \caption{Spectroscopic redshift validation of a cluster using {\it SPIDERS} follow-up observations, as described in Sect. ~\ref{sect:spiders}. The top-left panel shows the distribution of red-sequence galaxies in Xclass0842 as a function of their projected distance to the X-ray centroid and the velocity offset to the mean cluster redshift. Blue dashed lines indicate the standard deviation of the velocity distribution. The top-right panel shows the location of those members projected on the sky, with a similar colour-coding as in the previous panel. The lower-left histogram shows the redshift distribution of galaxies in the red sequence with an associated spectroscopic measurement. The lower-right histogram shows the distribution of members (in blue) and non-members (in grey).}
        \label{fig:spider_cluster}
\end{figure*}

\section{The X-CLASS cluster catalogue}

\subsection{Redshift validation}

\subsubsection{Visual inspection}

 Before the compilation of the X-CLASS public catalogue of galaxy clusters, which we present below, we undertook a thorough inspection of all sources considering the availability of wide and recent optical photometric and spectroscopic surveys. We most frequently used DECaLS \citep{Dey2019}, PanSTARRS \citep{Flewelling2020}, and SDSS-DR16 \citep{Blanton2017}, which provided deeper optical images than those of DSS used in our first pass, and a plethora of spectroscopic data. Our procedure included: combined visual inspection of the X-ray and optical images, matching of our candidates with previously released X-ray or optical catalogues of galaxy clusters, collecting all available spectroscopic data to confirm the redshift of the cluster, and producing all relevant meta-data for inclusion in the database.

\begin{table}
    \centering
    \caption{Status of the X-CLASS cluster candidates. Sources classified as `provisional' are sources that may not correspond to actual clusters.}
    \begin{tabular}{r|c|c|c|}
         status & Dec$>$0 & Dec$<$0 & total \\
         \hline
        confirmed & 556 & 426 & 982\\ 
        tentative & 63 & 31 & 94 \\
        photometric & 51 & 151 & 202 \\ 
        no status (no $z$) & 109  &  172  &  281 \\
        provisional (no $z$) &  48  &  39  &  87 \\
\hline
        all & {\bf 827} & {\bf 819 } & {\bf 1646 }
    \end{tabular}
    \label{tab:distr}
\end{table}

More specifically, our first action was to search for previously detected galaxy clusters that coincide with our X-ray detection. This information was mostly readily available from our previous matching, but nevertheless we used the NED database for our sources and have included all recent information. All relevant data are stored in the database and are available to the public. When available, the cross-matching provided a first estimation of the cluster redshift. Then, especially in the case where no matching cluster was found in the literature, we had to visually inspect the optical and the X-ray image of each cluster candidate. Our first choice for the optical band was PanSTARRS-DR1 colour images (from g and z bands) where we found that a concentration of red cluster galaxies is visible up to a redshift of $z\sim1$. PanSTARRS is available for the full northern sky and down to a declination of $-30$ degrees. For a much more limited sky area, DECaLS survey was also available, which is essential for high-redshift cluster candidates. For the rest of the clusters with no deep optical data, we used the DSS images, as we had previously done during our first pass. 

Then, for each cluster with an initial estimation for its redshift, we computed the projected 500 kpc radius. We selected this limiting radius as it roughly corresponds to an average $R_{500}$ radius for a moderately rich cluster, and we expect to locate the vast majority of cluster galaxies within this range. The same radius was used for the XXL survey cluster sample \citep{Adami2018}. We then collected data from NED and, when available, from SDSS-DR16, because recent spectroscopic data from the latest release were not yet available in NED. To validate the redshift of an X-ray cluster we chose to implement the same guidelines as for the XXL survey catalogue \citep{Adami2018}. Therefore, the redshift of a cluster can be categorised as follows:
\begin{itemize}
    \item {\bf Confirmed:} if three or more galaxies with concordant spectroscopic redshifts are found within the 500 kpc radius from the centre of the X-ray detection, or alternatively, if the spectrum of the brightest cluster galaxy (BCG) is available.
    \item {\bf Tentative:} if one or two galaxies with concordant spectroscopic redshifts are found within the 500 kpc radius.
    \item {\bf Photometric:} if only photometric redshift information is available in the literature or from our previous dedicated follow-up \citep{Ridl2017}.
    \item {\bf Provisional:} for cases where the available information does not allow us to verify the existence of a galaxy cluster in this position. Further follow-up observations are needed to safely classify these sources. Although these sources are part of the X-CLASS catalogue, they are not included in the online database, but can be found in Appendix D.     
\end{itemize}

All cluster candidates were reviewed by at least two researchers. However, visual inspection and redshift confirmation was especially critical for the 225 sources previously classified as `provisional', for which we followed the more thorough procedure described in the following section.  

\subsubsection{Provisional sources\label{sect:dubius}}

 As provisional cases were more challenging for the reviewers, each source was further inspected by two more consortium members. Their task was to simply classify the detection as a true or false cluster, considering all available optical and X-ray data. Out of the 225 sources, 70 were already validated as true clusters (49 confirmed, 12 tentative, and 9 photometric) and were used to test the reliability of the reviewers. For this step, we considered the classifications of seven researchers (different from those of the previous section), while we discarded those of the two researchers that gave the greatest number of false answers with respect to the already validated sources.   
 
Finally, we require full agreement between the researchers in order to characterise a source as a true cluster, or a spurious or misclassified detection, while sources with discrepant classifications kept their `provisional' status. More precisely, 33 provisional sources were classified as `inexistent' and were discarded from the catalogue, while 35 were identified as true clusters and entered the catalogue without any redshift information. The remaining 87 sources retained the `provisional' status, and, though kept in the main public catalogue, are not published online, but are included in Appendix D of the present paper for the interested reader, especially as this list may contain high-redshift cluster candidates.  

 Misclassified detections are usually due to a point-like source embedded in the extended X-ray emission of a nearby cluster, while high background may also play an important role. The majority of the spurious or misclassified cluster candidates have no visually detected counterpart in the optical band; their X-ray emission was limited within less than $10''$ radius and was usually centred on an optical point source. In addition, we applied machine-learning classification to our provisional sources in order to test the efficiency of such methods in difficult cases. The results are presented in Appendix A.

\subsubsection{Spectroscopic follow-up with SPIDERS\label{sect:spiders}}

A subsample of the previous X-CLASS sample was selected for spectroscopic follow-up within the frame of SPIDERS (SPectroscopic IDentification of {\it eROSITA} Sources), a dedicated survey for a homogeneous and complete sample of X-ray active galactic nuclei and galaxy clusters over a large fraction of the sky \citep{Clerc2016, Dwelly2017, Salvato2018, Comparat2020}. The (BOSS) spectrograph mounted on the SDSS-2.5m telescope at Apache Point Observatory \citep{Gunn2006} was used. The sample was compiled based on the correlation of X-ray sources from X-CLASS with the RedMapper optical cluster catalogue, as described in \citet{Sadibekova2014} and \citet{Clerc2016}. The current catalogue contains 124 validated clusters (out of the 142 targeted) with SPIDERS follow-up spectroscopy up to a redshift of $z\sim$0.6. The program led to the collection of 1134 spectra in X-CLASS red sequences, with a redshift success rate approaching 99\% \citep{Clerc2020}. The median number of galaxies with concordant redshifts used for the redshift validation was ten. Membership is assigned with an algorithm iteratively performing $3\sigma$ clipping around the mean redshift. Manual refinements are then allowed in the case of degenerate situations or failures due to the low number of spectroscopic redshifts available. An example cluster is shown in Fig.~\ref{fig:spider_cluster} \citep[see also][]{Kirkpatrick2021}. The list of clusters confirmed with SPIDERS spectroscopy can be found in Table \ref{table:spiders} of the Appendix.  

\subsection{The new X-CLASS catalogue}
\label{sect:countrate}

\begin{figure}
        \includegraphics[scale=0.55]{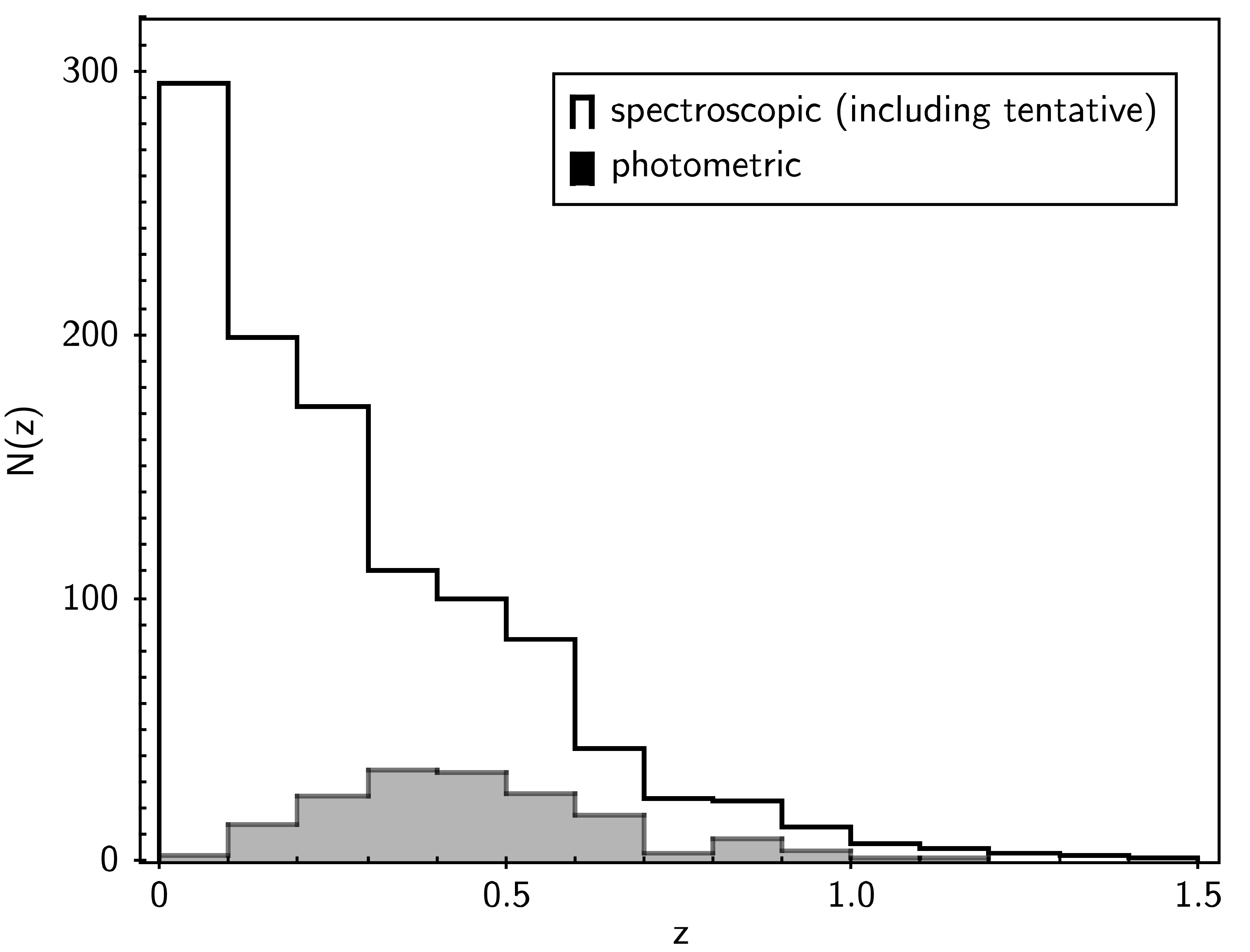}
                \includegraphics[scale=0.55]{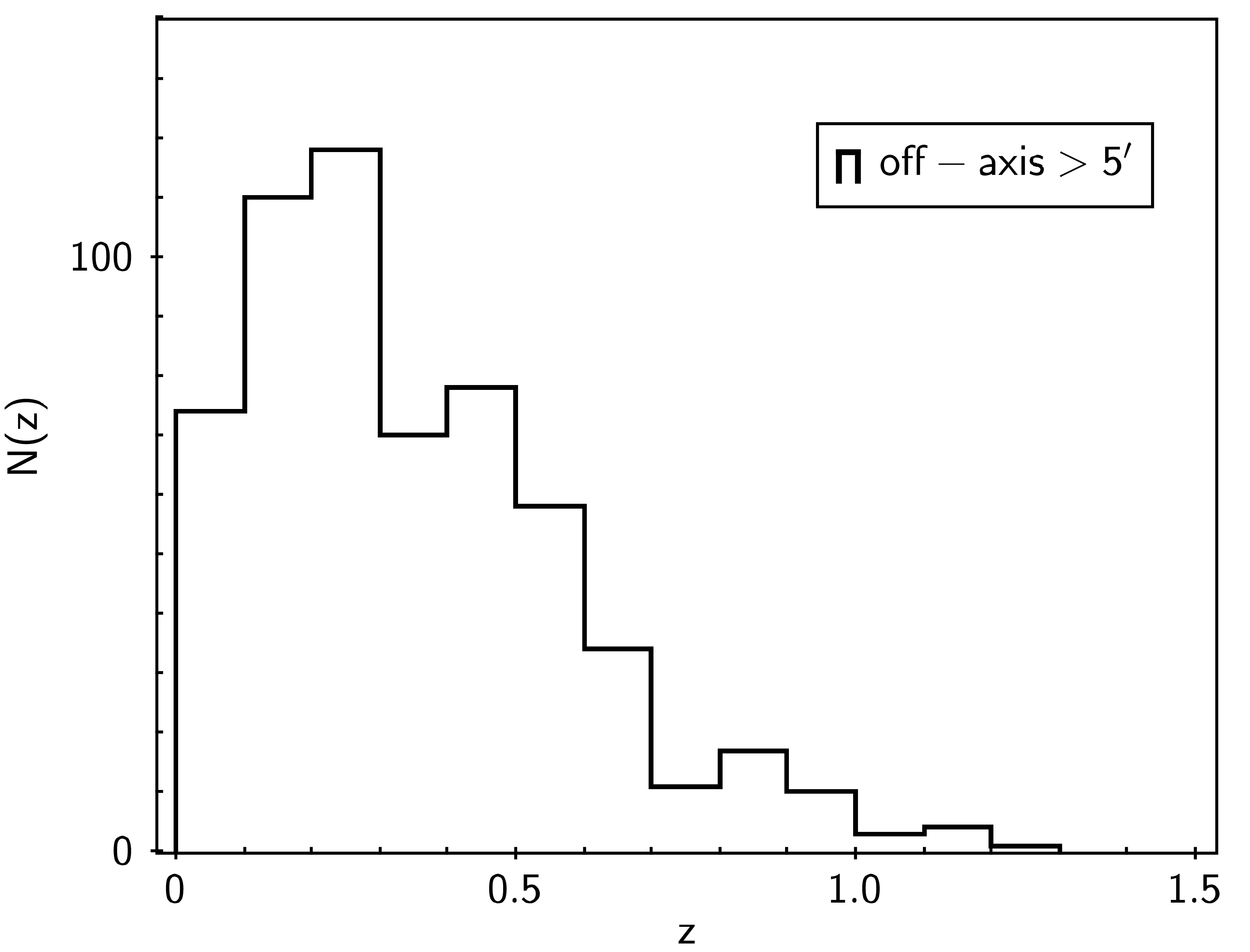}
        \caption{Top panel: Redshift distribution of the X-CLASS clusters with available spectroscopic or photometric redshift. Bottom panel: Redsift distribution for clusters found outside the central $5'$ of the XMM pointings, which excludes pointed observations that bias our sample towards low-redshift values. This distribution peaks around $z_{med}=0.29$.}
        \label{fig:zdistr}
\end{figure}

 Following the above classification scheme, the catalogue comprises 1278 X-ray-selected clusters with redshift information: 982 spectroscopically confirmed clusters, 94 with a tentative redshift, and 202 with a photometric redshift. These three categories represent $\sim$78\% of the total cluster catalogue and their redshift distribution is presented in Fig.~\ref{fig:zdistr}. The final catalogue includes an additional 281 clusters with no redshift information. The results are summarised in Table \ref{tab:distr} where they are also split between the two hemispheres, because the availability of observational data is very different. This leads to less confirmed clusters and more with no redshift information in the southern hemisphere. 

In addition to the above sources, a small number of detections were flagged as `inexistent' ($\sim$2\%) and were removed from the catalogue, while the 87 sources that were classified as `provisional' can be found in Table D.1 of the current paper. We caution the interested reader that this list may include a number of spurious or misclassified sources.   

Coordinates and redshifts of all cluster galaxies are stored in the cluster database. In the case of confirmed clusters, a histogram with all available spectroscopic redshifts within the corresponding search radius was also produced using the online tool in NED.

\subsection{X-ray count-rate measurements}

In the database we provide two different X-ray count-rate measurements: the pipeline quantities and a more accurate measurement using a count-rate curve of growth method developed by our team \citep{Clerc2012a}. 

The pipeline value is the total background-subtracted source count rate in the three XMM-Newton detectors (MOS1+MOS2+PN) in the $[0.5-2]$ keV band. This is an automatic measurement provided by the XAmin pipeline fitting procedure (raw pipeline output). It is a measure of the total count rate integrated out to infinity under the assumption that the best-fitting model is correct. The background is locally defined within a box around each source as a flat component that includes both the photon (vignetted) and the particle background, and is fitted simultaneously with the source model parameters. This is a simplified approach because the particle background is in general not flat across the EPIC detectors. Nevertheless, it is justified by the small extent of the vast majority of our sources.

\begin{figure}
\centering
        \includegraphics[scale=0.5,angle=0]{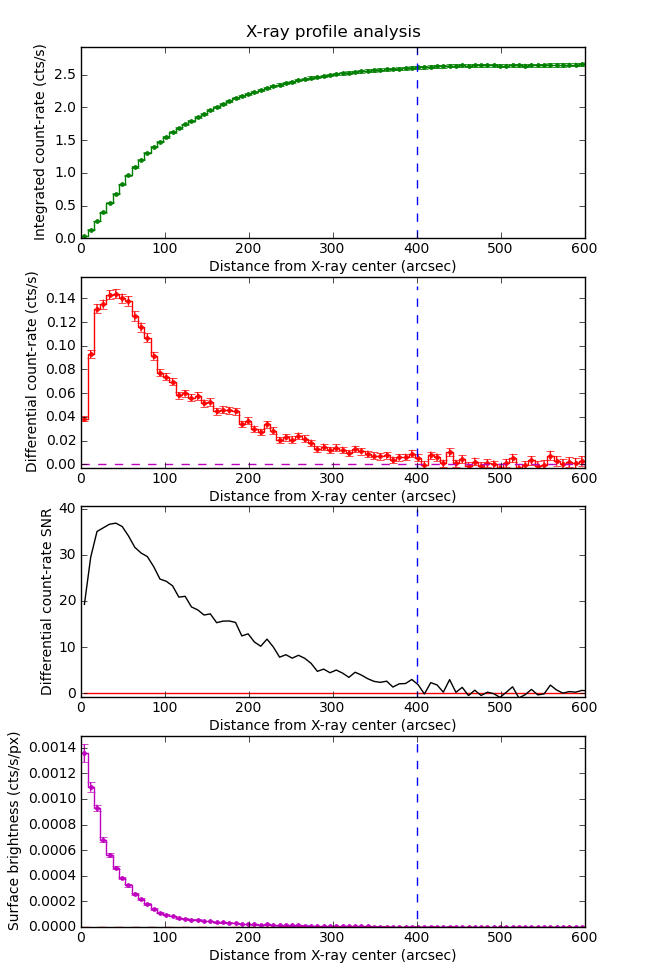}
        \caption{Interactive background-subtracted count-rate measurement of the X-ray cluster Xclass0047, as described in Sect \ref{sect:countrate}. The blue dashed line in each of the panels indicates the manually defined radius, within which the count-rate measurement is performed.}
        \label{fig:flux}
\end{figure} 

\begin{figure}
\centering
        \includegraphics[scale=0.5,angle=0]{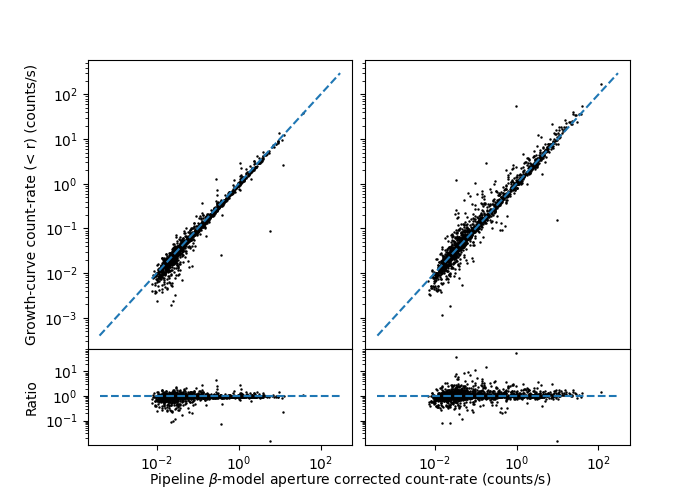}
        \caption{Comparison between the count rate measured by the source detection pipeline and the manual curve of growth analysis for the 1559 galaxy clusters in the sample. The values represent the equivalent on-axis count rate in the [0.5--2]\,keV energy band combining all three detectors, estimated within identical apertures $r$. In the left panel, $r=1$\,arcmin. In the right panel $r=R_{fit}$, a radius that is unique to each source and adapted to the signal-to-noise ratio in the measurement images. In both panels, the dashed line represents equality. Uncertainties are only available for manual measurements and are not shown in this figure.}
        \label{fig:compa_cr}
\end{figure} 

\begin{figure*}
        \includegraphics[width=\linewidth]{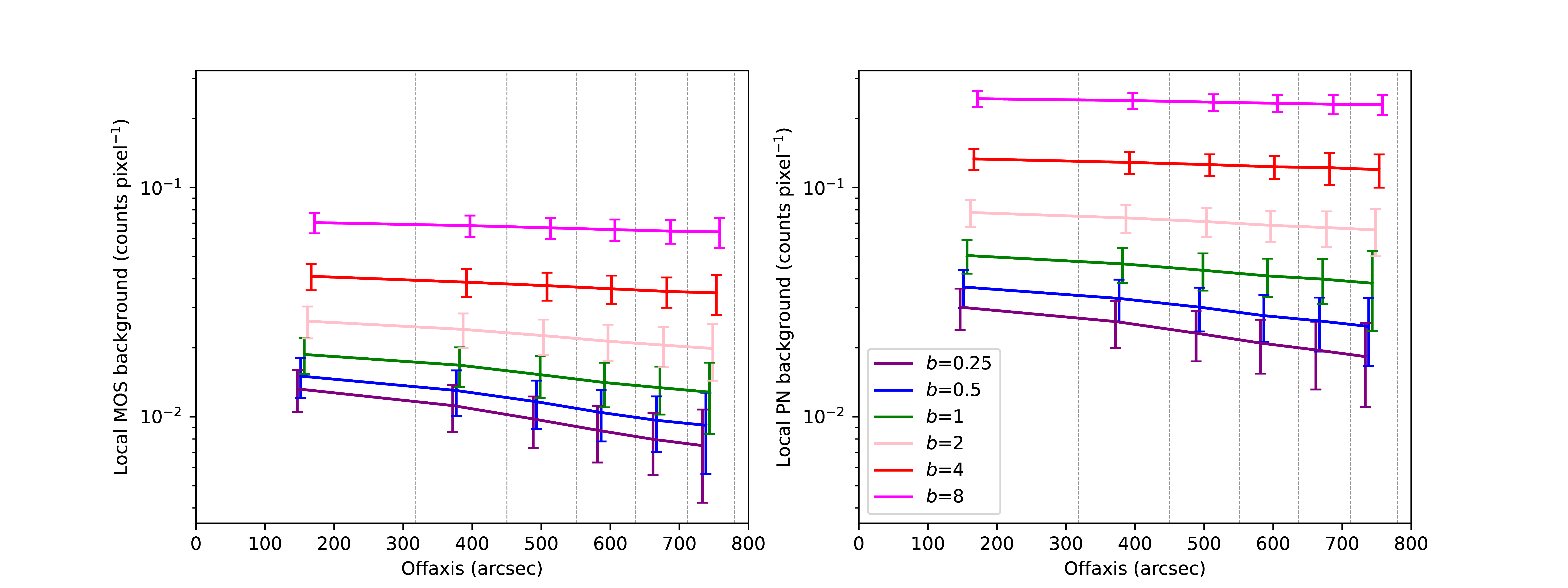}
        \caption{Calibration of the particle background $b$ parameter as a function of the local background fit by XAmin 3.5 in various equal-area off-axis bins for MOS (left) and PN (right panel) based on the point-source simulations. Off-axis bin boundaries are represented with vertical dashed lines and data points are slightly shifted on the x-axis for readability. Here, $b=1$ is the so-called nominal particle background, i.e. close to the most common level encountered in X-CLASS observations. In simulations, particle background is injected at a level of $b$ times the nominal value. The pixel size is $2.5''\times2.5''$.}
        \label{fig:back}
\end{figure*} 

In addition, all sources detected with the XAmin pipeline were subjected to a semi-interactive procedure, described in sect. 2.4 of \citetalias[][]{Clerc2012a}, in order to provide more accurate X-ray flux measurements. The sources were assumed to have a circular symmetry to extrapolate measurements on missing parts (masks, CCD gaps, and borders) of the detectors, and to integrate the count rate in concentric annuli. The possibility to interactively alter the segmentation masks was open to the users. The count rate $(cts/s)$ is the mean number of cluster X-ray photons collected by the three detectors in the direction of the optical axis in one second. The total count rate is the sum of all detector count-rate measurements. A count-rate growth curve as a function of cluster radius is computed from the total count rate. The background was modelled similarly to the respective XAmin automatic procedure. The flat and vignetted background levels were fitted on data extracted in a circular annulus around the source of interest, whose width and position are manually adjusted.
 
The procedure consists of two steps. The first is an interactive (manual) mode enabling the user to: (a) refine the X-ray cluster centre, (b) remove or correct areas incorrectly masked by XAmin (CCD gaps, unresolved blended sources, FOV edge cases), (c) re-estimate the background level according to the cluster brightness and extension to get a more precise count-rate measurement, (d) optimise the measurements in cases where the source is detected on the missing part of MOS1,
and (e) set a more accurate and reliable value for the source radii $R_{fit}$ when the growth curve algorithm has failed because of background overestimation (field source contamination, missing part of MOS1, edge effects). In usual conditions, $R_{fit}$ corresponds to the annulus in which the cluster count-rate uncertainty is compatible with background fluctuations.

The second step is automatically executed when the cluster parameters in the interactive mode have been set. During this step, the count rates are computed in six different bands, namely $[0.5-2]$, $[2-10]$, $[0.5-0.9]$, $[1.3-2]$, $[2-5]$ and $[5-7]$ keV, using a full exposure to obtain the highest signal-to-noise ratio. Settings and measurements for each cluster separately are available in the X-CLASS database on their $profile$ page. An example of the above procedure is illustrated in Fig.~\ref{fig:flux}.

A comparison between the two kinds of count-rate measurements is shown in Fig.~\ref{fig:compa_cr} for 1559 clusters for which both measurements are available. For the purpose of this comparison, the count-rate growth curve of each source is evaluated at a fixed angular aperture radius $r=60\arcsec$ and at the source radius $r=R_{fit}$. On the other hand, the pipeline count-rate measurement is aperture-corrected by means of the best-fit surface brightness model.
For sources brighter than 0.1\,counts\,s$^{-1}$ the agreement between the two measurements is very good: within arcminute-sized apertures, the manually measured count rate is 6\% lower on average than the pipeline count rate~; it is 4\% higher within $R_{fit}$. At lower count rates, manual measurements enable a more accurate centre positioning, refined background estimates, and more comprehensive source masking, hence recovering most of the failures due to automated model fitting.

\subsection{Cluster selection function}

In order to model the X-ray-extended selection function of the X-CLASS catalogue, we produced XMM simulations enriched with additional beta-models of a smaller extent, but otherwise identical to those used in \citetalias{Clerc2012a}. These simulations faithfully reproduce the characteristics of the detectors. We added AGNs following a published logN--logS relation \citep{Marconi03}, and unresolved AGNs modelled as a vignetted background component. Unvignetted particle background was also added, parameterised with a factor $b$. This factor represents the level of enhancement compared to the nominal background level, $b$=1. Simulations are performed with values of $b$=0.25, 0.5, 1, 2, 4, and 8, and the values of the added particle background as a function of off-axis distance are presented in Fig.~\ref{fig:back}. Optionally, extended sources as beta-models are distributed at random places on the detector. Their total count rate (in 0.5-2 keV) and apparent core radius are varied while beta is held fixed at 2/3. The range of count rates spans (0.0025 counts/s -- 0.5 counts/s). The range of core radii is (3, 5, 10, 20, 50, 100) arcsec.

The number of simulations 
amounts to 540 
for each value of $b$ and for exposure times of 10 and 20 ks. These were all processed with XAmin v3.5 as was briefly described in Sect. 2.1 to exactly mimic observations. We use a $37.5''$ matching radius for clusters, independently of their extent, flux, and so on, for consistency with previous works \citep[e.g.][]{pacaud2006}. Sources with multiple associations are ascribed to their nearest neighbour.
 
\begin{figure}
        \includegraphics[scale=0.45]{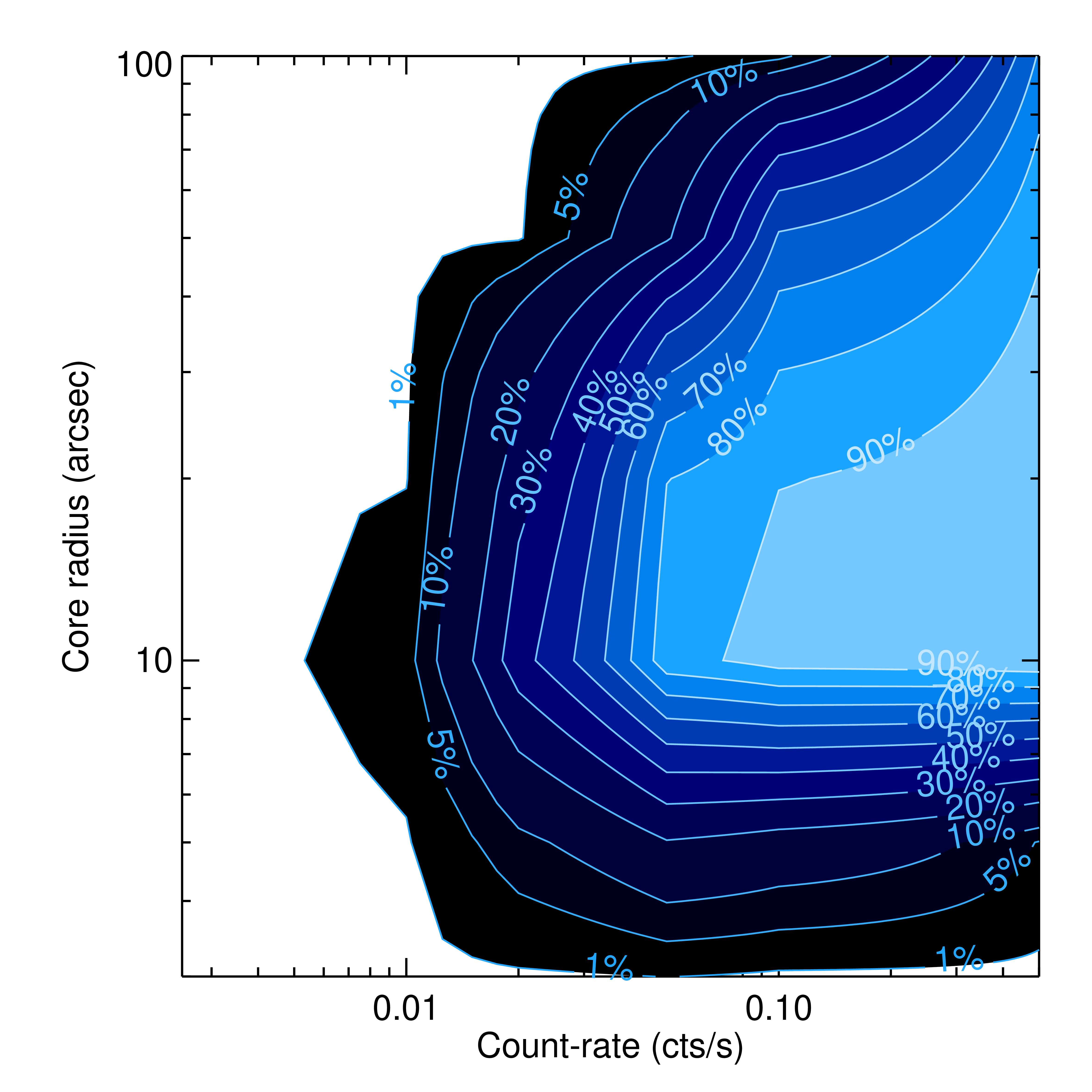}

        \caption{Contours of the C1 detection probability as a function of total count rate $[0.5-2]$ keV and input core radius of the beta model. Exposure time of Texp=10 ks and background value of $b=1$ were used.}
        \label{fig:Prob}
\end{figure}

\begin{figure}
    \includegraphics[width=\linewidth]{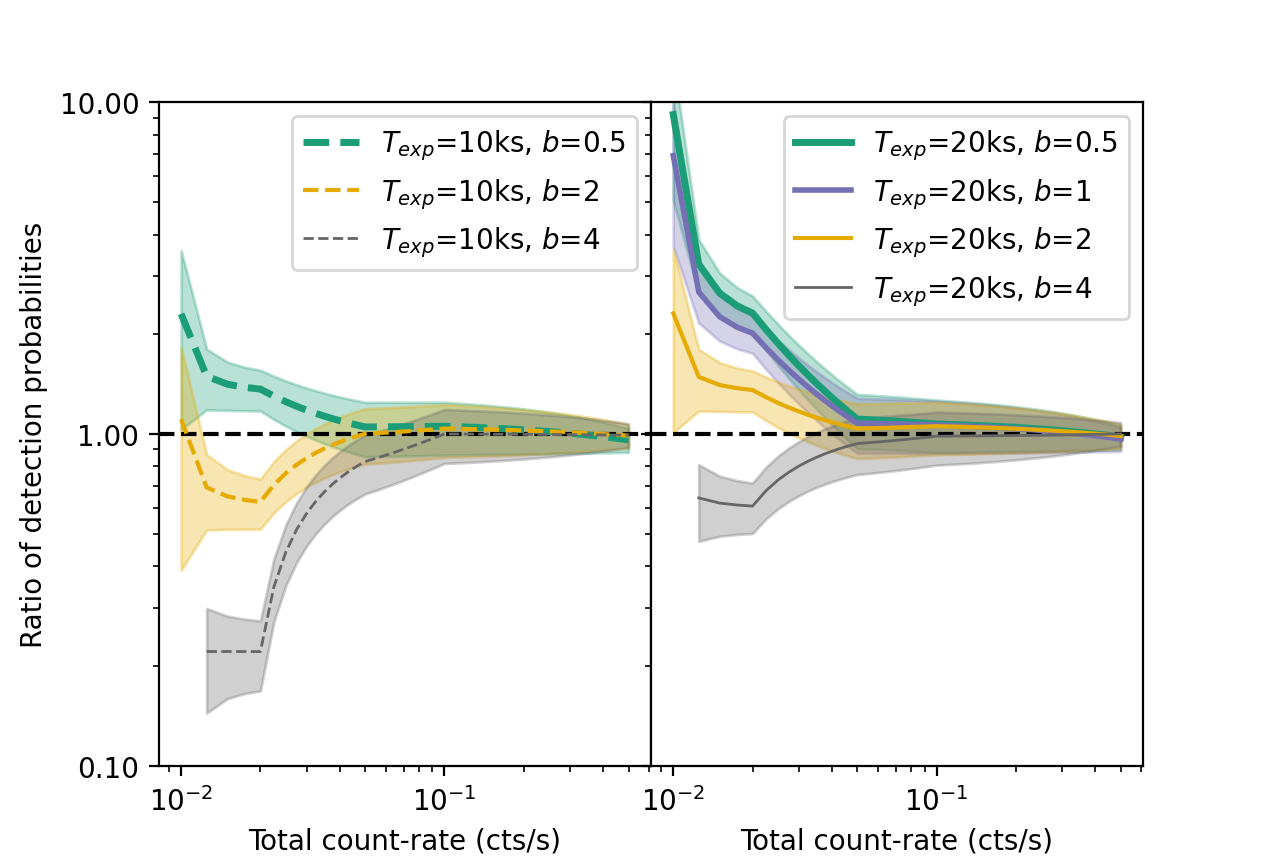}
        \caption{Variation of the C1 detection efficiency with exposure time and background level. Curves show the ratio relative to the probability shown in Fig.~\ref{fig:Prob} (exposure time of 10\,ks and $b=1$), assuming an apparent core radius $R_c=20\arcsec$. Shaded regions represent the standard deviation around these values, which depends on the number of simulated clusters. Only points with more than nine detected simulated clusters are shown.}
        \label{fig:Prob_rc20}
\end{figure}

In Fig.~\ref{fig:Prob} we present the detection probability of sources within $13'$ off-axis radius from the centre of the observation, for an exposure time of Texp=10 ks and for background level $b$=1 . It is apparent that the X-CLASS catalogue is not a flux-limited sample, but the selection is rather two-dimensional depending on the extent of the source. Figure~\ref{fig:Prob_rc20} illustrates the impact of increasing the background level or the exposure time on the selection of typical X-CLASS clusters (core radii around 20 arcsec). A twofold increase in exposure time provides more numerous low-flux clusters as long as the particle background level remains below three times the fiducial value $b=1$.

\begin{landscape}
\centering
\begin{table}
\caption{X-CLASS cluster database view.\newline (1) X-CLASS unique ID number and link to images and pipeline outputs, (2) right ascension (pipeline automatic measurement), (3) declination (pipeline automatic measurement), (4) right ascension (interactive refined measurement$^{\star}$), (3) declination (interactive refined measurement$^{\star}$), (6) link to NED data, (7) observation ID and link to the observation and data processing information page, (8) link to the redshift information page, (9) redshift status, (10) total X-ray count-rate in the [0.5--2 keV] band (pipeline output), (11) link to the interactive X-ray count-rate measurement page.\newline$^{\star}$for more details see sect. \ref{sect:countrate}}
\begin{center}
\begin{tabular}{|c|r|r|r|r|c|c|c|c|c|c|}
\hline    
Xclass& \multicolumn{1}{|c|}{RA pipeline} & \multicolumn{1}{|c|}{Dec pipeline} & \multicolumn{1}{|c|}{RA measured} & \multicolumn{1}{|c|}{Dec measured} & NED & Obs  & redshift & status  & total rate & profile \\
 & \multicolumn{1}{|c|}{(deg.)} & \multicolumn{1}{|c|}{(deg.)} & \multicolumn{1}{|c|}{(deg.)} & \multicolumn{1}{|c|}{(deg.)} &  &   &  &   & (counts/sec) & \\
 (1) & \multicolumn{1}{|c|}{(2)} & \multicolumn{1}{|c|}{(3)} & \multicolumn{1}{|c|}{(4)} & \multicolumn{1}{|c|}{(5)} & (6) & (7)  & (8) & (9)  & (10) & (11)\\
\hline
&&&&&&&&&&\\
{\color{blue}\underline {0020}} & 193.4380      &        10.1954        &        193.4380        &        10.1951        &
        {\color{blue}\underline {go}}   &       {\color{blue}\underline {0001930301\_10ks}}             &       {\color{blue}\underline {0.654}}        &       confirmed       &       0.049   &       {\color{blue}\underline {data}} \\
{\color{blue}\underline {0023}} & 194.2860      &        -17.4119       &        194.2920        &        -17.4064       &
        {\color{blue}\underline {go}}   &       {\color{blue}\underline {0010420201\_10ks}}             &       {\color{blue}\underline {0.047}}        &       confirmed       &       3.738   &       {\color{blue}\underline {data}} \\
{\color{blue}\underline {0033}} & 193.6790      &        -29.2227       &        193.6740        &        -29.2230       &
        {\color{blue}\underline {go}}   &       {\color{blue}\underline {0030140101\_10ks}}     &       {\color{blue}\underline {0.056}}        &       confirmed       &       5.882   &       {\color{blue}\underline {data}}         \\
{\color{blue}\underline {0034}} & 193.5950      &        -29.0162       &        193.5930        &        -29.0131       &
        {\color{blue}\underline {go}}   &       {\color{blue}\underline {0030140101\_10ks}}             &       {\color{blue}\underline {0.053}}        &       confirmed       &       4.362   &       {\color{blue}\underline {data}} \\
{\color{blue}\underline {0035}} & 196.2740      &        -10.2802       &        196.2740        &        -10.2787       &
        {\color{blue}\underline {go}}   &       {\color{blue}\underline {0032141201\_10ks}}             &       {\color{blue}\underline {0.34}} &       photometric     &       0.047   &       {\color{blue}\underline {data}} \\
{\color{blue}\underline {0038}} & 36.5674       &        -2.6651        &        36.5677 &        -2.6663        &
        {\color{blue}\underline {go}}   &       {\color{blue}\underline {0037981801\_10ks}}             &       {\color{blue}\underline {0.056} }&      confirmed       &       0.167   &       {\color{blue}\underline {data}} \\
{\color{blue}\underline {0039}} & 36.4987       &        -2.8272        &        36.4990 &        -2.8275        &
        {\color{blue}\underline {go}}   &       {\color{blue}\underline {0037981801\_10ks}}             &       {\color{blue}\underline {0.281} }&      confirmed       &       0.033   &       {\color{blue}\underline {data}} \\
{\color{blue}\underline {0040}} & 35.1871       &        -3.4339        &        35.1886 &        -3.4339        &
        {\color{blue}\underline {go}}   &       {\color{blue}\underline {0037982601\_10ks}}             &       {\color{blue}\underline {0.327} }&      confirmed       &       0.050   &       {\color{blue}\underline {data}} \\
{\color{blue}\underline {0042}} & 150.1230      &        -19.6282       &        150.1220        &        -19.6292       & 
        {\color{blue}\underline {go}}   &       {\color{blue}\underline {0041180301\_10ks}}     &               &       no redshift        & 0.057         &       {\color{blue}\underline {data}} \\
{\color{blue}\underline {0044}} & 202.4460      &        11.6835        &        202.4490        &        11.6848        &
    {\color{blue}\underline {go}}       &       {\color{blue}\underline {0041180801\_10ks}}     &       {\color{blue}\underline {0.204} }&      confirmed       &       0.087   &       {\color{blue}\underline {data}}         \\
{\color{blue}\underline {0047}} & 172.9830      &        -19.9229       &        172.9800        &        -19.9271       &
        {\color{blue}\underline {go}}   &       {\color{blue}\underline {0042341001\_10ks}}     &       {\color{blue}\underline {0.307} }&      confirmed       &       3.254   &       {\color{blue}\underline {data}}                 \\
{\color{blue}\underline {0048}} & 173.0280      &        -19.8611       &        173.0280        &        -19.8614       & 
        {\color{blue}\underline {go}}   &       {\color{blue}\underline {0042341001\_10ks}}             &       {\color{blue}\underline {0.307}}        &       confirmed       &       0.154   &       {\color{blue}\underline {data}}                 \\
{\color{blue}\underline {0050}} & 172.8110      &        -19.9326       &        172.8130        &        -19.9343       &
        {\color{blue}\underline {go}}   &       {\color{blue}\underline {0042341001\_10ks}}     &       {\color{blue}\underline {0.46}} &       photometric     &       0.025   &       {\color{blue}\underline {data}}         \\
{\color{blue}\underline {0051}} & 177.6130      &        1.7580 &        177.6160        &        1.7580 & 
        {\color{blue}\underline {go}}   &       {\color{blue}\underline {0044740201\_10ks}}     &               &       no redshift        & 0.036 &       {\color{blue}\underline {data}}         \\
{\color{blue}\underline {0054}} & 145.9370      &        16.7402        &        145.9380        &        16.7381        &
        {\color{blue}\underline {go}}   &       {\color{blue}\underline {0046940401\_10ks}}             &       {\color{blue}\underline {0.18}} &       confirmed       &       0.136   &       {\color{blue}\underline {data}}         \\
{\color{blue}\underline {0056}} & 145.8820      &        16.6656        &        145.8860        &        16.6671        &
        {\color{blue}\underline {go}}   &       {\color{blue}\underline {0046940401\_10ks}}             &       {\color{blue}\underline {0.255}}        &       confirmed       &       0.202   &       {\color{blue}\underline {data}}         \\
{\color{blue}\underline {0057}} & 145.9920      &        16.6871        &        145.9950        &        16.6875        &
        {\color{blue}\underline {go}}   &       {\color{blue}\underline {0046940401\_10ks}}             &       {\color{blue}\underline {0.253}}        &       confirmed       &       0.057   &       {\color{blue}\underline {data}}         \\
{\color{blue}\underline {0059}} & 31.9565       &        2.1553 &        31.9576 &        2.1567 &
        {\color{blue}\underline {go}}   &       {\color{blue}\underline {0052140301\_20ks}}     &       {\color{blue}\underline {0.334}}        &       photometric     &       0.042   &       {\color{blue}\underline {data}}         \\
{\color{blue}\underline {0062}} & 44.1414       &        0.1037 &        44.1417 &        0.1033 &
        {\color{blue}\underline {go}}   &       {\color{blue}\underline {0056020301\_10ks}}             &       {\color{blue}\underline {0.362}}        &       confirmed       &       0.778   &       {\color{blue}\underline {data}}                 \\
{\color{blue}\underline {0065}} & 339.2510      &        -15.2730       &        339.2520        &        -15.2731       &
        {\color{blue}\underline {go}}   &       {\color{blue}\underline {0056021601\_10ks}}             &       {\color{blue}\underline{0.31}}  &       photometric     &       0.316   &       {\color{blue}\underline {data}} \\
{\color{blue}\underline {0075}}&        10.4501 &-9.4575&       10.4507&        -9.4569&{\color{blue}\underline {go}}&{\color{blue}\underline{0723802201\_20ks}}&       {\color{blue}\underline{0.056}}&        confirmed&15.529&       {\color{blue}\underline {data}} \\

{\color{blue}\underline {0078}}&        10.7223 &-9.5697&       10.7225 &-9.5701&{\color{blue}\underline {go}}&{\color{blue}\underline{0065140201\_10ks}}&{\color{blue}\underline{       0.41}}& photometric&0.101&      {\color{blue}\underline {data}} \\

{\color{blue}\underline {0079}}&        10.5228 &-9.6026&       10.5231 &-9.6029&{\color{blue}\underline {go}}&{\color{blue}\underline{0065140201\_10ks}}&{\color{blue}\underline{       0.055}}&        tentative&0.033&        {\color{blue}\underline {data}} \\

{\color{blue}\underline {0082}}&        39.4926 &-52.3934&      39.4929&        -52.3937&{\color{blue}\underline {go}}&{\color{blue}\underline{0067190101\_10ks}}&{\color{blue}\underline{       0.136}}&        confirmed&0.259&        {\color{blue}\underline {data}} \\

{\color{blue}\underline {0083}}&        148.424 &1.6999&        148.424 &1.6995 &{\color{blue}\underline {go}}&{\color{blue}\underline{0070940401\_10ks}}&{\color{blue}\underline{0.097}}&       confirmed&0.796&        {\color{blue}\underline {data}} \\

{\color{blue}\underline {0086}}&        348.765 &-58.9351&      348.766 &-58.9354&{\color{blue}\underline {go}}&{\color{blue}\underline{0081340301\_10ks}}&       {\color{blue}\underline{0.44}}& photometric&0.034&      {\color{blue}\underline {data}} \\

{\color{blue}\underline {0087}}&        349.094&        -59.0752&       349.095&        -59.0756&{\color{blue}\underline {go}}&{\color{blue}\underline{0081340301\_10ks}}&{\color{blue}\underline{       0.62}}& photometric&0.036&      {\color{blue}\underline {data}} \\

{\color{blue}\underline {0088}}&        183.394 &2.8953&        183.395 &2.8963&{\color{blue}\underline {go}}&  {\color{blue}\underline{0081340801\_10ks}}&{\color{blue}\underline{0.410}}&     confirmed&0.190&        {\color{blue}\underline {data}} \\

{\color{blue}\underline {0095}}&        190.102 &-11.8008&      190.103&        -11.8011&{\color{blue}\underline {go}}&{\color{blue}\underline{0084030101\_10ks}}&       {\color{blue}\underline{0.19}}& photometric&0.078&      {\color{blue}\underline {data}} \\

{\color{blue}\underline {0096}}&        9.2778  &9.1566 &9.2765&        9.1583& {\color{blue}\underline {go}}&{\color{blue}\underline{0084230201\_20ks}}& {\color{blue}\underline{0.252}}&      confirmed&2.441&{\color{blue}   {\color{blue}\underline {data}}}        \\

&&&&&&&&&&\\
\hline

\end{tabular}
\end{center}
\label{table:1}
\end{table}
\end{landscape}

\section{The X-CLASS database\label{sect:database}}
The X-CLASS database is built using the MySQL database management system and is accessible to a user through a Java Web Application. The database contains all catalogue meta-data, images and plots, which are available in PNG format generated from pipeline output. The cluster table is dynamically generated using a selection interface and is then displayed as an HTML web page. Several criteria (redshift, X-CLASS name, x-ray parameters, sky area, etc.) can be applied to refine (constrain) the cluster selection. For each cluster position, there are pre-configured requests to the external astronomical data servers (CDS, NED, NASA/IPAC IRSA) enabled to display optical counterparts and overlays within $3~arcmin$. A user manual along with relevant documentation is also available online, as well as the list of X-CLASS publications. The database is hosted in CC-IN2P3 (Centre de Calcul de l'IN2P3 at Lyon in France) and is publicly available online in {\href{https://xmm-xclass.in2p3.fr}{\color{blue} https://xmm-xclass.in2p3.fr/}}.  

\subsection{Cluster table\label{sect:cluster_table}}
The first 20 entries are listed in Table \ref{table:1}, sorted according to increasing RA, while the full version of the catalogue table can be retrieved from the VizieR server at the CDS (\textcolor{red}{link will be added}). From the database graphical interface, the catalogue table provides the following fields  for each
cluster:
\begin{enumerate}
\item $Xclass$ $-$ a unique cluster identifier. 
\item $RA$  $-$ pipeline-measured right ascension. 
\item $Dec$ $-$ pipeline-measured declination. 
\item $Obs$ $-$ XMM pointing where the cluster is detected. 
\item $NED$ $-$ the NED search is within $3~arcmin$, the page is automatically generated using the AladinLite webapp. 
\item $redshift$ $-$ cluster redshift, linked to the redshift validation page.
\item $status$ $-$ redshift validation status (e.g. photometric, spectroscopic).
\item $total~rate$ $-$ pipeline total count rate.  
\item $profile$ $-$ more accurate count-rate quantities from {\em FluxMes} measurements for six X-ray bands within $R_{fit}$
\end{enumerate}

\section{Discussion and comparison with other catalogues}

\begin{figure}
\centering
        \includegraphics[scale=0.55,angle=0]{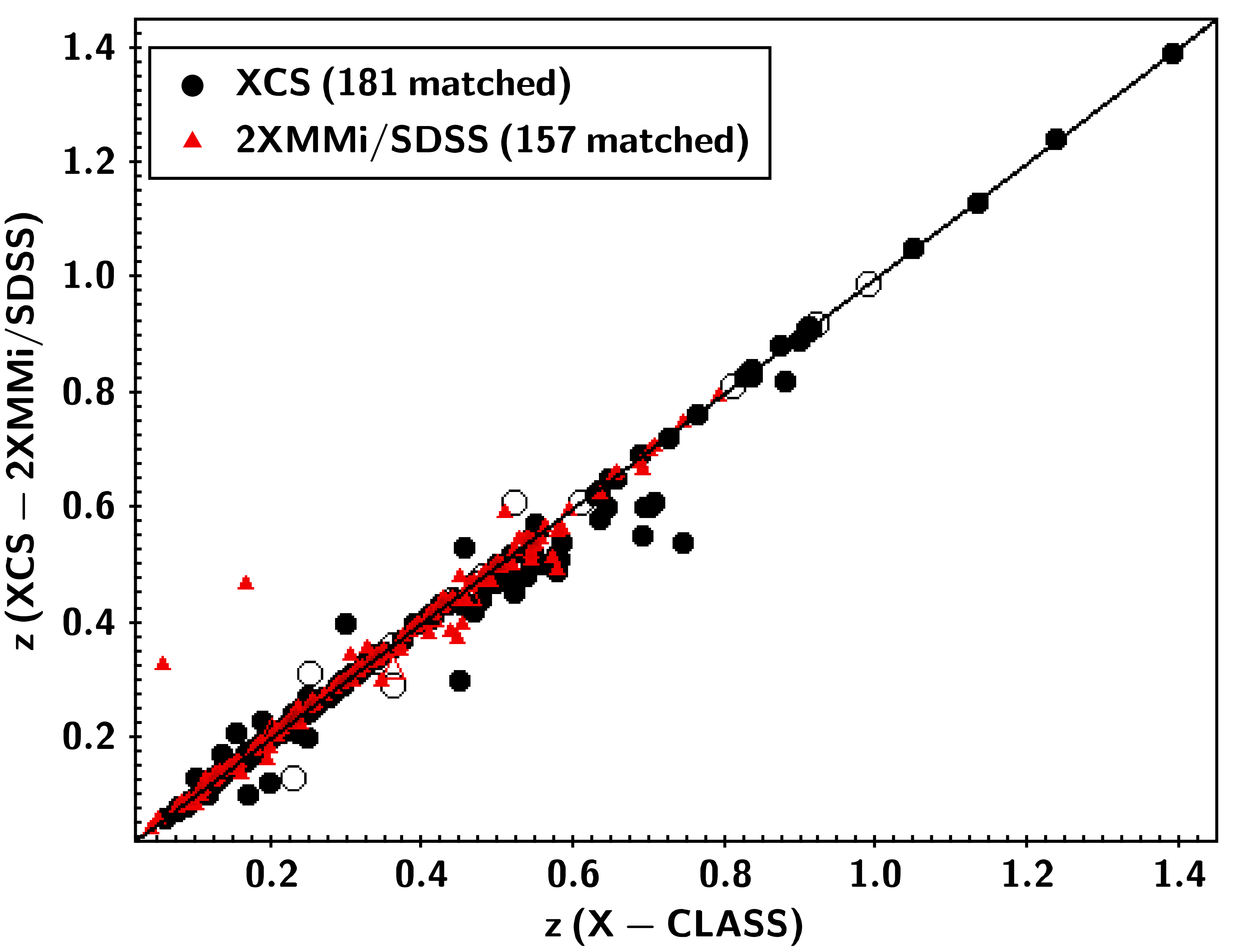}
        \includegraphics[scale=0.55,angle=0]{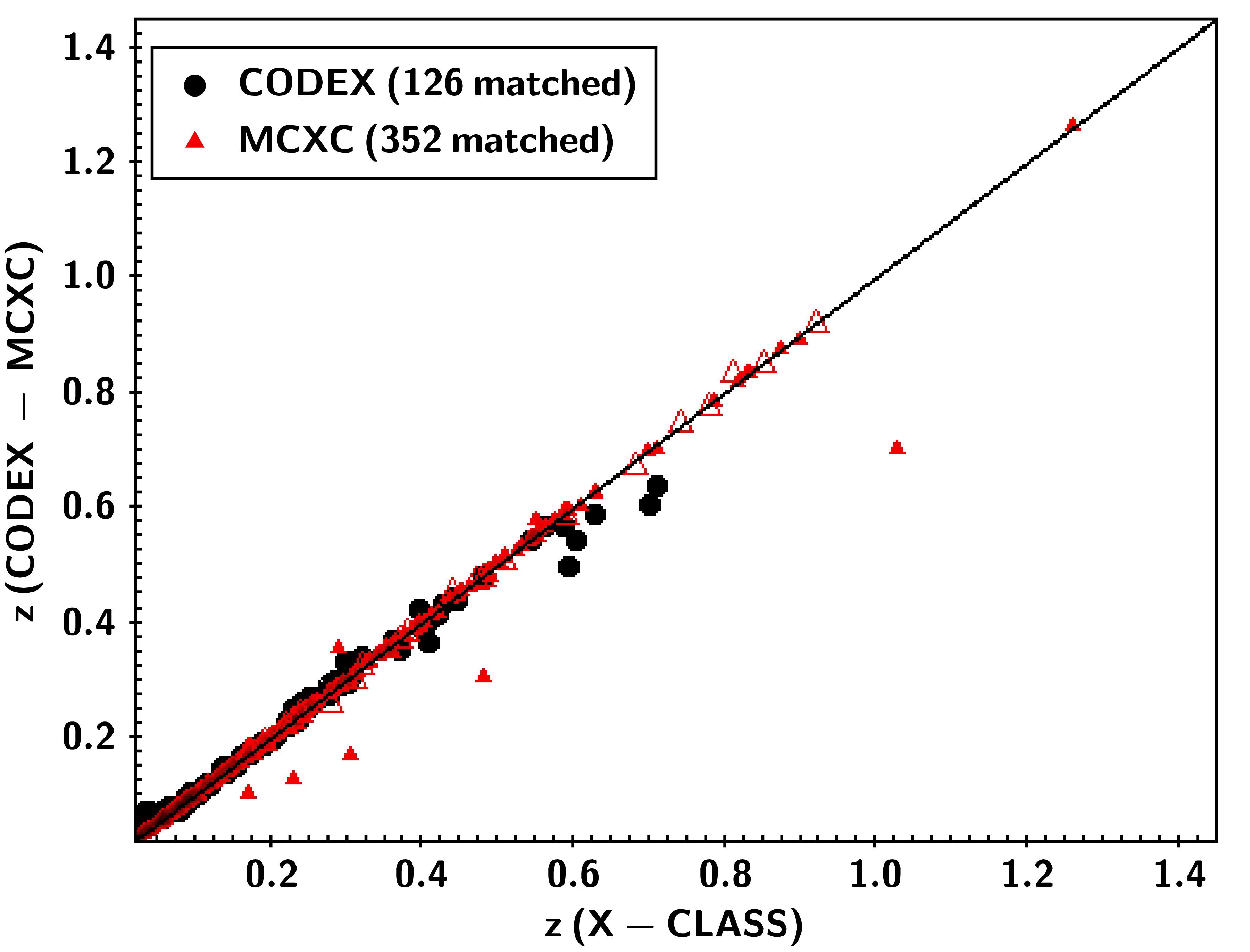}
        \caption{Redshift comparison between the X-CLASS catalogue and various other X-ray-selected cluster samples. All catalogues include both clusters with spectroscopic and photometric redshifts, except for CODEX, which includes only photometric redshift. The filled (open) shapes mark clusters with spectroscopic (photometric) redshift.}
        \label{fig:red_comp}
\end{figure}

On one hand, the homogeneous selection of the X-CLASS clusters on homogenised observations of 10 and 20 ks exposures simplifies the computation of their selection function and their use in cosmology studies. On the other hand, because of the above selection, the comparison with other similar datasets is difficult \citepalias[see also relevant discussion in][]{Clerc2012a}. Nevertheless, within $1'$  we cross-correlated the X-CLASS catalogue with other X-ray-selected cluster catalogues in order to examine the redshift agreement of the common sources. In Fig.~\ref{fig:red_comp} we present such a comparison with two catalogues of X-ray-detected clusters by XMM-Newton, XCS \citep{Mehrtens2012}, and 2XMMi/SDSS \citep{Takey2014}, and two by ROSAT, MCXC \citep{Piffaretti2011} and CODEX \citep{Finoguenov2020}. There is good agreement between redshifts, especially with the MCXC catalogue. We note that in many cases redshifts were obtained from the same source. In cases of large discrepancies we examined our cluster identifications and redshift validation. In most of the discrepant cases, our more recent and updated data allow for a more accurate redshift estimation, while in a limited number of cases, the matched X-ray detections are correlated with different optical counterparts.      

\begin{figure}
\centering
        \includegraphics[scale=0.55,angle=0]{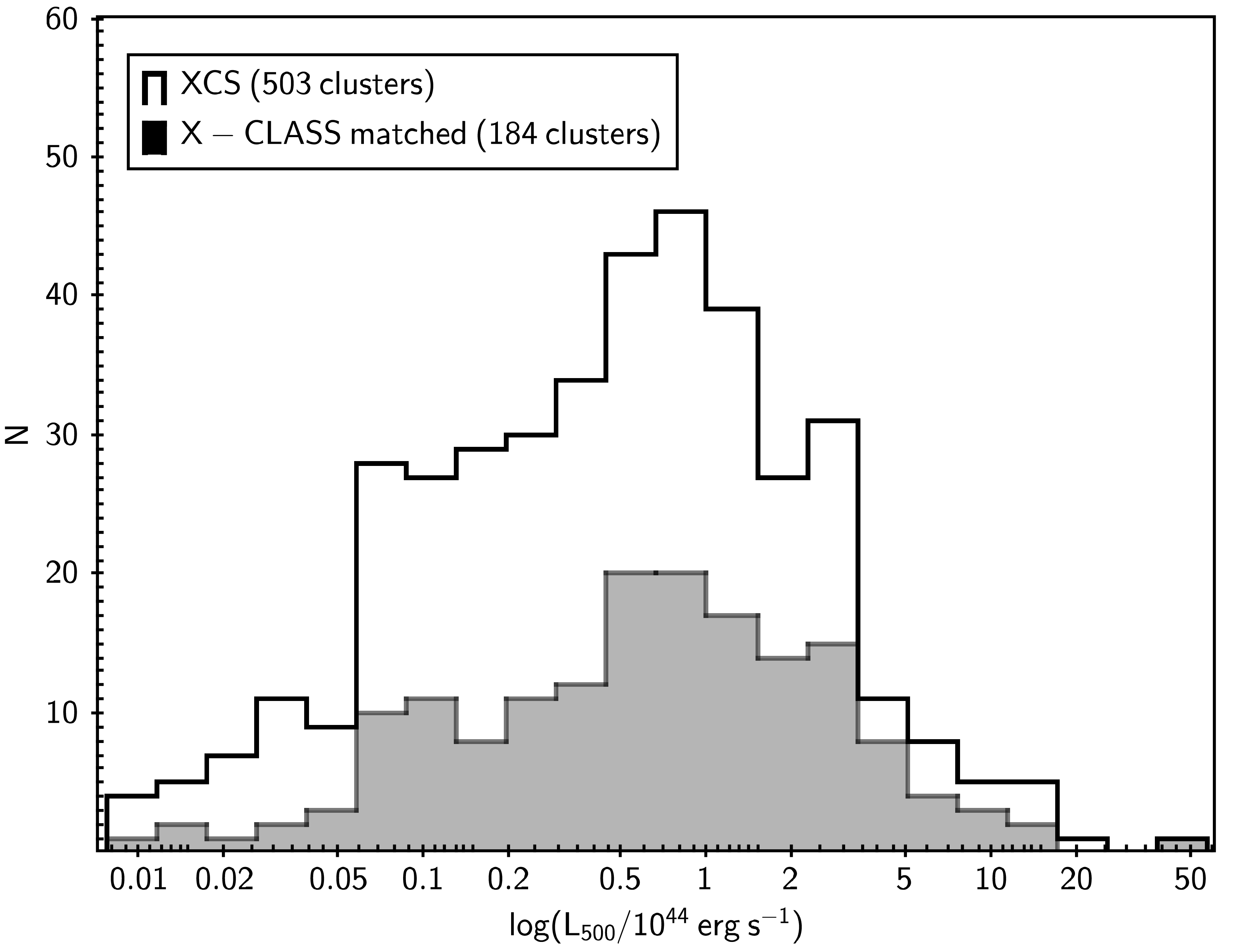}
        \caption{X-ray bolometric luminosity $L_{500}[0.05-100]$ keV distribution of the matched X-CLASS and XCS cluster catalogues. According to a KS two-sample test the null hypothesis that the two samples originate from the same parent distribution cannot be rejected.}
        \label{fig:lum500}
\end{figure}

We also examined the X-ray luminosity distribution of the matched X-CLASS/XCS cluster sample in order to identify any systematic bias in our selection. In Fig. \ref{fig:lum500} we plot the distribution of $L_{500}[0.05-100]$ keV, as estimated in \citet{Mehrtens2012} within the $R_{500}$ radius, for both the matched and the full XCS sample. The two samples have very similar luminosity distributions, as also confirmed by a two-sample Kolmogorov-Smirnov test. Therefore, the matched sample is an unbiased subsample of the total XCS sample.

In addition we cross-correlated our catalogue with the 4XMM-DR10 catalogue \citep{Webb2020}. This catalogue includes a large number of extended X-ray sources and their fitting parameters, but does not include any redshift estimation or further examination of their nature. The large majority of the X-CLASS sources are correlated to an extended source in the 4XMM catalogue within $1'$ radius. However, 191 sources are absent, most of which have a provisional status or no redshift information, and inspection of their optical counterpart places them at high redshift. In addition, they usually have a low value for the extent likelihood parameter, as computed by the XAmin pipeline. Nevertheless, a high fraction ($\sim$27\%)  of the uncorrelated clusters have a `confirmed' or `photometric' status, usually below $z=0.6$.

\section{Summary}

We present the detection pipeline, selection function, visual inspection, screening, and redshift confirmation of a large number of X-ray-detected galaxy clusters in 4176 archived {\it XMM-Newton} images. The total number reaches 1646 clusters over 269 deg$^2$ and the catalogue is publicly accessible via an interactive database constructed and maintained by the X-CLASS collaboration. The database not only allows the selection of a cluster subsample based on a large variety of criteria, but also gives access to a wealth of meta-data, images, plots, and supplementary information.    

The homogeneous selection of the cluster samples over 10 and 20 ks exposures allows the use of the X-CLASS catalogue for cosmological analyses, provided the impact of pointed clusters is accounted for. In addition, the large sky area makes it suitable as a test bed for current and future large-area cluster surveys, such as the ones that will be carried out by the {\it eROSITA} \citep{Merloni2012,Predehl2021}, {\it Athena} \citep{Nandra2013}, and {\it Euclid} \citep{Laureijs2011} missions, considering the large amount of human effort and interaction required for the compilation of the present catalogue, which could not be extended to the huge datasets of these missions.      

\begin{acknowledgements}
The Saclay team acknowledges long term support from the Centre National d’Etudes Spatiales. 

This research has made use of "Aladin sky atlas" developed at CDS, Strasbourg Observatory, France (\citep{aladinCDS} and \citep{2014ASPC..485..277B}). The cross-matching
has made using of the NASA/IPAC Extragalactic Database (NED),
which is operated by the Jet Propulsion Laboratory, California Institute of Technology,
under contract with the National Aeronautics and Space Administration.

The Pan-STARRS1 Surveys (PS1) and the PS1 public science archive have been made possible through contributions by the Institute for Astronomy, the University of Hawaii, the Pan-STARRS Project Office, the Max-Planck Society and its participating institutes, the Max Planck Institute for Astronomy, Heidelberg and the Max Planck Institute for Extraterrestrial Physics, Garching, The Johns Hopkins University, Durham University, the University of Edinburgh, the Queen's University Belfast, the Harvard-Smithsonian Center for Astrophysics, the Las Cumbres Observatory Global Telescope Network Incorporated, the National Central University of Taiwan, the Space Telescope Science Institute, the National Aeronautics and Space Administration under Grant No. NNX08AR22G issued through the Planetary Science Division of the NASA Science Mission Directorate, the National Science Foundation Grant No. AST-1238877, the University of Maryland, Eotvos Lorand University (ELTE), the Los Alamos National Laboratory, and the Gordon and Betty Moore Foundation.

Funding for the Sloan Digital Sky Survey IV has been provided by the Alfred P. Sloan Foundation, the U.S. Department of Energy Office of Science, and the Participating Institutions. SDSS acknowledges support and resources from the Center for High-Performance Computing at the University of Utah. The SDSS web site is www.sdss.org.

SDSS is managed by the Astrophysical Research Consortium for the Participating Institutions of the SDSS Collaboration including the Brazilian Participation Group, the Carnegie Institution for Science, Carnegie Mellon University, Center for Astrophysics / Harvard \& Smithsonian (CfA), the Chilean Participation Group, the French Participation Group, Instituto de Astrofísica de Canarias, The Johns Hopkins University, Kavli Institute for the Physics and Mathematics of the Universe (IPMU) / University of Tokyo, the Korean Participation Group, Lawrence Berkeley National Laboratory, Leibniz Institut für Astrophysik Potsdam (AIP), Max-Planck-Institut für Astronomie (MPIA Heidelberg), Max-Planck-Institut für Astrophysik (MPA Garching), Max-Planck-Institut für Extraterrestrische Physik (MPE), National Astronomical Observatories of China, New Mexico State University, New York University, University of Notre Dame, Observatório Nacional / MCTI, The Ohio State University, Pennsylvania State University, Shanghai Astronomical Observatory, United Kingdom Participation Group, Universidad Nacional Autónoma de México, University of Arizona, University of Colorado Boulder, University of Oxford, University of Portsmouth, University of Utah, University of Virginia, University of Washington, University of Wisconsin, Vanderbilt University, and Yale University.

The Legacy Surveys consist of three individual and complementary projects: the Dark Energy Camera Legacy Survey (DECaLS; NSF's OIR Lab Proposal ID 2014B-0404; PIs: David Schlegel and Arjun Dey), the Beijing-Arizona Sky Survey (BASS; NSF's OIR Lab Proposal ID 2015A-0801; PIs: Zhou Xu and Xiaohui Fan), and the Mayall z-band Legacy Survey (MzLS; NSF's OIR Lab Proposal ID 2016A-0453; PI: Arjun Dey). DECaLS, BASS and MzLS together include data obtained, respectively, at the Blanco telescope, Cerro Tololo Inter-American Observatory, The NSF's National Optical-Infrared Astronomy Research Laboratory (NSF's OIR Lab); the Bok telescope, Steward Observatory, University of Arizona; and the Mayall telescope, Kitt Peak National Observatory, NSF's OIR Lab. The Legacy Surveys project is honored to be permitted to conduct astronomical research on Iolkam Du'ag (Kitt Peak), a mountain with particular significance to the Tohono O'odham Nation.

The NSF's OIR Lab is operated by the Association of Universities for Research in Astronomy (AURA) under a cooperative agreement with the National Science Foundation.

This project used data obtained with the Dark Energy Camera (DECam), which was constructed by the Dark Energy Survey (DES) collaboration. Funding for the DES Projects has been provided by the U.S. Department of Energy, the U.S. National Science Foundation, the Ministry of Science and Education of Spain, the Science and Technology Facilities Council of the United Kingdom, the Higher Education Funding Council for England, the National Center for Supercomputing Applications at the University of Illinois at Urbana-Champaign, the Kavli Institute of Cosmological Physics at the University of Chicago, Center for Cosmology and Astro-Particle Physics at the Ohio State University, the Mitchell Institute for Fundamental Physics and Astronomy at Texas A\&M University, Financiadora de Estudos e Projetos, Fundacao Carlos Chagas Filho de Amparo, Financiadora de Estudos e Projetos, Fundacao Carlos Chagas Filho de Amparo a Pesquisa do Estado do Rio de Janeiro, Conselho Nacional de Desenvolvimento Cientifico e Tecnologico and the Ministerio da Ciencia, Tecnologia e Inovacao, the Deutsche Forschungsgemeinschaft and the Collaborating Institutions in the Dark Energy Survey. The Collaborating Institutions are Argonne National Laboratory, the University of California at Santa Cruz, the University of Cambridge, Centro de Investigaciones Energeticas, Medioambientales y Tecnologicas-Madrid, the University of Chicago, University College London, the DES-Brazil Consortium, the University of Edinburgh, the Eidgenossische Technische Hochschule (ETH) Zurich, Fermi National Accelerator Laboratory, the University of Illinois at Urbana-Champaign, the Institut de Ciencies de l'Espai (IEEC/CSIC), the Institut de Fisica d'Altes Energies, Lawrence Berkeley National Laboratory, the Ludwig-Maximilians Universitat Munchen and the associated Excellence Cluster Universe, the University of Michigan, the National Optical Astronomy Observatory, the University of Nottingham, the Ohio State University, the University of Pennsylvania, the University of Portsmouth, SLAC National Accelerator Laboratory, Stanford University, the University of Sussex, and Texas A\&M University.

BASS is a key project of the Telescope Access Program (TAP), which has been funded by the National Astronomical Observatories of China, the Chinese Academy of Sciences (the Strategic Priority Research Program "The Emergence of Cosmological Structures" Grant XDB09000000), and the Special Fund for Astronomy from the Ministry of Finance. The BASS is also supported by the External Cooperation Program of Chinese Academy of Sciences (Grant 114A11KYSB20160057), and Chinese National Natural Science Foundation (Grant 11433005).

The Legacy Survey team makes use of data products from the Near-Earth Object Wide-field Infrared Survey Explorer (NEOWISE), which is a project of the Jet Propulsion Laboratory/California Institute of Technology. NEOWISE is funded by the National Aeronautics and Space Administration.

The Legacy Surveys imaging of the DESI footprint is supported by the Director, Office of Science, Office of High Energy Physics of the U.S. Department of Energy under Contract No. DE-AC02-05CH1123, by the National Energy Research Scientific Computing Center, a DOE Office of Science User Facility under the same contract; and by the U.S. National Science Foundation, Division of Astronomical Sciences under Contract No. AST-0950945 to NOAO.

\end{acknowledgements}

\bibliographystyle{aa} 
\bibliography{citations} 

\begin{appendix}

\section{Convolutional neural network classification of the provisional sources.}

Convolutional neural networks (CNNs) are state-of-the art machine-learning tools for image classification. In \citet{Kosiba2020}, our team developed a custom CNN architecture for automatic classification of galaxy cluster candidates into two classes, `galaxy cluster' and `non-cluster'. To train the network, we provided it with X-ray and optical images of approximately\,1500 galaxy cluster candidates together with their manual classifications as given by experts from the X-CLASS collaboration. The network used those data to learn how to understand patterns of different classes of objects and projection and instrumental effects of the data. When successfully trained, we evaluated the performance of the network  on a sample of 85 spectroscopically confirmed galaxy clusters and 85 objects we classified as non-clusters. Our network achieved $\sim$\,90\,$\%$ accuracy. For more technical details on the construction of the data and our CNN architecture, we kindly refer the interested reader to \citet{Kosiba2020}.

\begin{figure}[ht]
        \includegraphics[scale=0.35,angle=270]{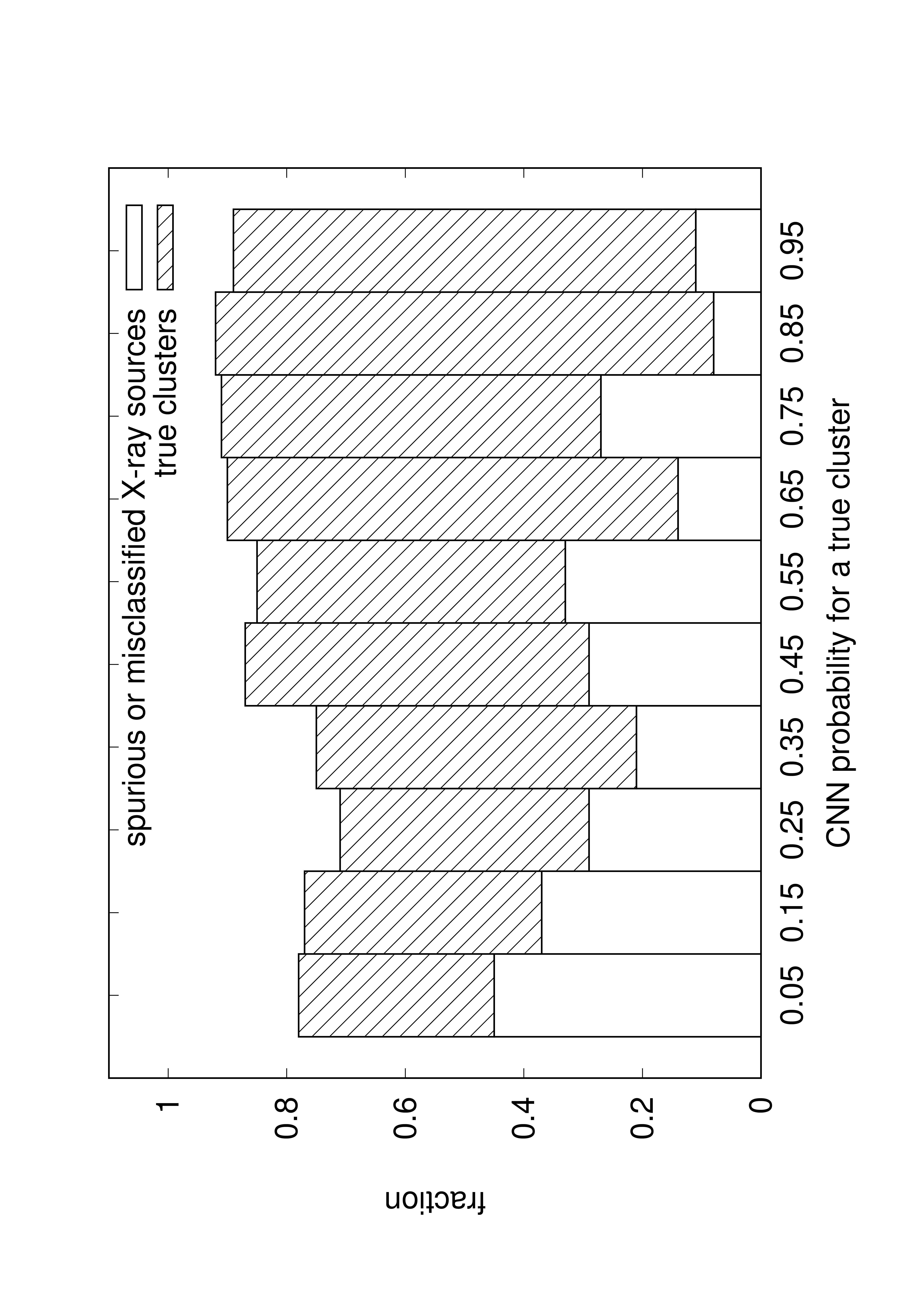}
        \caption{Agreement between CNN classification and expert opinion. The hatched area shows the fraction of sources agreed upon by all researchers as being true clusters, while the empty area shows the false detections agreed upon by all researchers. The percentage of sources for which the  expert opinions are mixed is omitted for clarity, but these sources account for the remaining fraction up to 1 for each bin. The number of sources in each bin varies from 10 to 33. CNNs are shown to be more effective in identifying true clusters in the last two bins (CNN prob.$>$80\%) and false detections in the first bin (CNN prob.$<$10\%).}
        \label{fig:CNN}
\end{figure}

Here, we compare the CNN classifications with the expert opinions on the
225 `provisional' sources as described in Sect. \ref{sect:dubius}. The results are illustrated in Fig. A.1. Mixed expert opinions are not included in the histogram but they can easily be inferred
as the remaining percentage in each bin. We clearly see that the CNN
method is most successful in classifying sources as clusters with
more than 80\% probability (the last two bins). The agreement
in both these bins is $\sim$\,80\%. We conclude that, as expected, automatic classification of X-ray sources is more efficient at identifying true bonafide clusters than at discarding spurious or misclassified ones.

We stress that the CNN was not trained on provisional sources because these could not be labelled as `galaxy clusters' or `non-clusters' by definition of this class. The provisional sources are galaxy cluster candidates that the experts were not sure how to classify in the first manual screening, making them the hardest-to-classify galaxy cluster candidates.

However, further training of our CNN on a large sample of difficult cases with known true classification (e.g. spectroscopic confirmation) would make it more reliable for classification of difficult provisional sources and especially useful for large-area surveys where human interaction will be impossible. 

\section{Detected sources on observed fields with less than three detectors}

In this section, we provide a table of the additional nine X-ray extended sources that were detected on observations with one or two missing detectors.

\begin{table}[ht]
    \centering
    \caption{X-ray clusters detected on XMM-Newton observations with less than three detectors. Redshift and status as defined in Sect. \ref{sect:cluster_table}}
    \begin{tabular}{|r|r|c|c|c|}
    \hline
          \multicolumn{1}{|c|}{RA} & \multicolumn{1}{|c|}{Dec} & redshift & status & available\\
\multicolumn{1}{|c|}{(deg.)} & \multicolumn{1}{|c|}{(deg.)}&&&detectors\\
\hline
         7.436  &       4.873   &       0.206   &        confirmed & MOS1+MOS2  \\
        8.664   &       -12.12  &       0.44    &        photometric & MOS1+MOS2        \\
        110.22  &       71.151  &       0.231   &        confirmed & MOS1+MOS2  \\
        131.797 &       34.813  &       0.552   &        confirmed & MOS1+MOS2  \\
        154.798 &       45.047  &       $-$     &       $-$      & MOS1+MOS2    \\
        155.151 &       45.005  &       $-$     &        provisional & MOS1+MOS2        \\
        175.353 &       -12.279 &       0.115   &        tentative & MOS1+MOS2  \\
        226.499 &       1.697   &       0.237   &        tentative &    PN \\
        156.059 &   4.193   &   -       &    provisional & PN \\
\hline
    \end{tabular}
    \label{tab:discarded}
\end{table}

\section{X-CLASS clusters spectroscopically confirmed with SPIDERS}

In Table \ref{table:spiders} we present the subsample of 124 clusters selected for spectroscopic follow-up within the frame of SPIDERS (see Sect. \ref{sect:spiders}). 
\onecolumn
\begin{table*}
\centering
\caption{X-CLASS clusters spectroscopically confirmed with SPIDERS (see Sect.~\ref{sect:spiders}). This table lists the systems that are validated with $N_{\rm mem}$ spectroscopic redshifts selected among $N_{\rm z}$ redshifts available in their red sequence. The observational status indicates `complete' if all selected targets led to a spectrum acquisition. The spectroscopic redshift uncertainty reflects the spread in the $N_{\rm mem}$ redshift values. A few systems marked with $^{(1)}$ were confirmed with only two spectroscopic members, hence no uncertainty is given on the redshift.}
\begin{tabular}{|l|c|c|c|c||l|c|c|c|c|}
\hline    
Xclass& $N_{\rm z}$ & Obs. status & $N_{\rm mem}$ & Zspec & Xclass& $N_{\rm z}$ & Obs. status & $N_{\rm mem}$ & Zspec\\
\hline
&&&&&&&&&\\
0039    &       12      &       complete        &       7       &       $0.2810 \pm 0.0009$     &       1624    &       15      &       complete        &       12      &       $0.2276 \pm 0.0009$     \\
0040    &       15      &       complete        &       12      &       $0.3274 \pm 0.0009$     &       1626    &       5       &       complete        &       5       &       $0.548 \pm 0.002$      \\
0062    &       18      &       complete        &       18      &       $0.362 \pm 0.002$      &       1627    &       16      &       complete        &       15      &       $0.3297 \pm 0.0009$     \\
0096    &       17      &       complete        &       16      &       $0.252 \pm 0.001$      &       1635    &       11      &       complete        &       8       &       $0.428 \pm 0.001$      \\
0099    &       26      &       complete        &       22      &       $0.2311 \pm 0.0007$     &       1637    &       24      &       complete        &       22      &       $0.206 \pm 0.001$      \\
0102    &       15      &       complete        &       14      &       $0.0593 \pm 0.0003$     &       1642    &       14      &       complete        &       9       &       $0.55 \pm 0.01$       \\
0103    &       12      &       complete        &       11      &       $0.1320 \pm 0.0005$     &       1674    &       17      &       complete        &       8       &       $0.580 \pm 0.001$      \\
0108    &       19      &       complete        &       14      &       $0.1949 \pm 0.0006$     &       1676    &       27      &       complete        &       14      &       $0.2881 \pm 0.0004$     \\
0109    &       12      &       complete        &       9       &       $0.478 \pm 0.001$      &       1678    &       24      &       complete        &       18      &       $0.409 \pm 0.002$      \\
0110    &       24      &       complete        &       14      &       $0.2703 \pm 0.0009$     &       1680    &       4       &       complete        &       3       &       $0.552 \pm 0.004$      \\
0169    &       10      &       complete        &       9       &       $0.320 \pm 0.002$      &       1686    &       11      &       complete        &       6       &       $0.3076 \pm 0.0009$     \\
0224    &       15      &       complete        &       11      &       $0.1423 \pm 0.0006$     &       1706    &       9       &       incomplete      &       6       &       $0.3321 \pm 0.0004$     \\
0245    &       10      &       complete        &       7       &       $0.1603 \pm 0.0003$     &       1737    &       21      &       complete        &       20      &       $0.276 \pm 0.001$      \\
0270    &       15      &       complete        &       12      &       $0.2452 \pm 0.0006$     &       1738    &       20      &       complete        &       18      &       $0.280 \pm 0.002$      \\
0336    &       17      &       complete        &       13      &       $0.421 \pm 0.001$      &       1758    &       7       &       complete        &       6       &       $0.342 \pm 0.001$      \\
0342    &       4       &       incomplete      &       3       &       $0.230 \pm 0.003$      &       1763    &       9       &       complete        &       4       &       $0.3124 \pm 0.0006$     \\
0343    &       17      &       incomplete      &       15      &       $0.351 \pm 0.001$      &       1764    &       15      &       complete        &       8       &       $0.311 \pm 0.001$      \\
0344    &       14      &       complete        &       11      &       $0.2909 \pm 0.0006$     &       1789    &       10      &       complete        &       2       &       $0.59^{(1)}$    \\
0347    &       3       &       incomplete      &       2       &       $0.26^{(1)}$    &       1807    &       16      &       complete        &       10      &       $0.4999 \pm 0.0006$     \\
0349    &       28      &       complete        &       22      &       $0.1537 \pm 0.0005$     &       1816    &       10      &       complete        &       3       &       $0.579 \pm 0.001$      \\
0361    &       8       &       complete        &       7       &       $0.0454 \pm 0.0007$     &       1817    &       9       &       complete        &       5       &       $0.579 \pm 0.002$      \\
0377    &       9       &       incomplete      &       8       &       $0.395 \pm 0.002$      &       1853    &       33      &       complete        &       26      &       $0.2972 \pm 0.0008$     \\
0574    &       8       &       complete        &       4       &       $0.497 \pm 0.002$      &       1854    &       14      &       complete        &       8       &       $0.519 \pm 0.001$      \\
0578    &       30      &       complete        &       29      &       $0.1396 \pm 0.0007$     &       1855    &       11      &       complete        &       8       &       $0.1865 \pm 0.0007$     \\
0615    &       11      &       complete        &       6       &       $0.255 \pm 0.001$      &       1866    &       2       &       complete        &       2       &       $0.44^{(1)}$    \\
0628    &       7       &       complete        &       6       &       $0.2323 \pm 0.0006$     &       1900    &       13      &       complete        &       8       &       $0.413 \pm 0.002$      \\
0630    &       5       &       complete        &       4       &       $0.373 \pm 0.001$      &       1904    &       34      &       complete        &       32      &       $0.0895 \pm 0.0003$     \\
0632    &       19      &       complete        &       14      &       $0.3947 \pm 0.0005$     &       1941    &       17      &       complete        &       14      &       $0.0996 \pm 0.0003$     \\
0638    &       6       &       complete        &       4       &       $0.500 \pm 0.001$      &       1957    &       23      &       complete        &       23      &       $0.2696 \pm 0.0009$     \\
0686    &       8       &       complete        &       4       &       $0.460 \pm 0.003$      &       1982    &       10      &       complete        &       6       &       $0.248 \pm 0.001$      \\
0706    &       5       &       complete        &       4       &       $0.609 \pm 0.002$      &       1983    &       10      &       complete        &       9       &       $0.3453 \pm 0.0004$     \\
0734    &       6       &       complete        &       4       &       $0.427 \pm 0.004$      &       2003    &       7       &       complete        &       5       &       $0.532 \pm 0.003$      \\
0740    &       14      &       complete        &       9       &       $0.3388 \pm 0.0008$     &       2026    &       8       &       complete        &       4       &       $0.527 \pm 0.001$      \\
0755    &       18      &       complete        &       14      &       $0.1969 \pm 0.0004$     &       2034    &       6       &       incomplete      &       5       &       $0.248 \pm 0.001$      \\
0841    &       13      &       complete        &       7       &       $0.551 \pm 0.003$      &       2036    &       13      &       complete        &       6       &       $0.328 \pm 0.001$      \\
0842    &       29      &       complete        &       22      &       $0.3009 \pm 0.0009$     &       2051    &       17      &       complete        &       9       &       $0.411 \pm 0.002$      \\
0890    &       19      &       complete        &       13      &       $0.3397 \pm 0.0007$     &       2080    &       11      &       complete        &       6       &       $0.427 \pm 0.002$      \\
0908    &       16      &       complete        &       11      &       $0.2744 \pm 0.0005$     &       2081    &       21      &       complete        &       19      &       $0.293 \pm 0.002$      \\
0953    &       12      &       complete        &       7       &       $0.1327 \pm 0.0002$     &       2088    &       13      &       complete        &       11      &       $0.0900 \pm 0.0009$     \\
0963    &       13      &       complete        &       12      &       $0.2477 \pm 0.0005$     &       2090    &       26      &       complete        &       25      &       $0.0904 \pm 0.0006$     \\
1013    &       10      &       complete        &       8       &       $0.4149 \pm 0.0005$     &       2093    &       21      &       complete        &       16      &       $0.2970 \pm 0.0008$     \\
1059    &       26      &       complete        &       22      &       $0.2791 \pm 0.0005$     &       2097    &       19      &       complete        &       18      &       $0.1125 \pm 0.0006$     \\
1062    &       26      &       complete        &       22      &       $0.1238 \pm 0.0004$     &       2109    &       25      &       complete        &       20      &       $0.2128 \pm 0.0006$     \\
1069    &       9       &       complete        &       9       &       $0.1328 \pm 0.0004$     &       2154    &       8       &       complete        &       7       &       $0.329 \pm 0.001$      \\
1086    &       7       &       complete        &       7       &       $0.423 \pm 0.002$      &       2155    &       9       &       complete        &       8       &       $0.390 \pm 0.001$      \\
1159    &       22      &       complete        &       20      &       $0.412 \pm 0.001$      &       2182    &       6       &       complete        &       6       &       $0.520 \pm 0.002$      \\
1185    &       11      &       complete        &       5       &       $0.492 \pm 0.002$      &       2208    &       13      &       complete        &       8       &       $0.1050 \pm 0.0004$     \\
1288    &       18      &       complete        &       13      &       $0.532 \pm 0.002$      &       2214    &       20      &       complete        &       15      &       $0.3004 \pm 0.0004$     \\
1307    &       28      &       complete        &       26      &       $0.0593 \pm 0.0004$     &       2272    &       13      &       complete        &       9       &       $0.254 \pm 0.003$      \\
1350    &       4       &       incomplete      &       3       &       $0.396 \pm 0.001$      &       2295    &       28      &       complete        &       22      &       $0.3695 \pm 0.0007$     \\
1351    &       11      &       incomplete      &       6       &       $0.543 \pm 0.002$      &       2328    &       2       &       complete        &       2       &       $0.45^{(1)}$    \\
1368    &       23      &       complete        &       20      &       $0.288 \pm 0.001$      &       2338    &       31      &       complete        &       29      &       $0.1406 \pm 0.0004$     \\
1369    &       12      &       complete        &       5       &       $0.284 \pm 0.002$      &       2340    &       12      &       complete        &       11      &       $0.443 \pm 0.002$      \\
1386    &       13      &       complete        &       12      &       $0.314 \pm 0.001$      &       2344    &       21      &       complete        &       21      &       $0.232 \pm 0.002$ \\   
1439    &       23      &       complete        &       21      &       $0.0573 \pm 0.0006$     &       2345    &       22      &       complete        &       20      &       $0.219 \pm 0.001$      \\
1443    &       31      &       complete        &       24      &       $0.0543 \pm 0.0006$     &       2347    &       28      &       complete        &       26      &       $0.1672 \pm 0.0007$     \\
1451    &       8       &       complete        &       6       &       $0.478 \pm 0.004$      &       2348    &       26      &       complete        &       25      &       $0.1921 \pm 0.0009$     \\
1452    &       7       &       complete        &       5       &       $0.447 \pm 0.004$      &       2350    &       23      &       complete        &       20      &       $0.412 \pm 0.001$      \\
1543    &       10      &       complete        &       10      &       $0.370 \pm 0.001$      &       2353    &       25      &       complete        &       23      &       $0.190 \pm 0.001$      \\
1544    &       19      &       complete        &       18      &       $0.370 \pm 0.001$      &       2358    &       30      &       complete        &       25      &       $0.0949 \pm 0.0007$     \\
1548    &       5       &       complete        &       5       &       $0.3281 \pm 0.0006$     &       2360    &       14      &       complete        &       10      &       $0.235 \pm 0.002$      \\
1622    &       21      &       complete        &       14      &       $0.0794 \pm 0.0003$     &       2363    &       14      &       complete        &       8       &       $0.312 \pm 0.002$      \\
\hline
\end{tabular}
    \label{table:spiders}
\end{table*}

\twocolumn

\section{Provisional sources}
Table \ref{table:prov}  presents the subsample of 87 candidate clusters for which the visual screening (see Sect. \ref{sect:dubius}) did not provide a conclusive result. These sources are not included in the online public catalogue.

\begin{table}[h!]
\centering
\caption{X-CLASS clusters classified as `provisional'.}
\begin{tabular}{|r|r|r|c||r|r|r|c|}
\hline    
Xclass& \multicolumn{1}{|c|}{RA} & \multicolumn{1}{|c|}{Dec} & total rate & Xclass& \multicolumn{1}{|c|}{RA} & \multicolumn{1}{|c|}{Dec} & total rate  \\
&\multicolumn{1}{|c|}{(deg.)} & \multicolumn{1}{|c|}{(deg.)} & (counts/sec) & & \multicolumn{1}{|c|}{(deg.)} & \multicolumn{1}{|c|}{(deg.)} & (counts/sec)\\
\hline
\hline
&&&&&&&\\
20070   &       0.792   &       -29.968 &       0.05688 &       3424    &       169.214 &       17.987  &       0.03811 \\
0489    &       11.785  &       25.278  &       3.66885 &       1894    &       169.572 &       7.971   &       0.01281 \\
0486    &       11.858  &       -25.126 &       0.03257 &       0231    &       172.902 &       -34.695 &       0.02231 \\
0508    &       16.507  &       -80.151 &       0.03299 &       2268    &       178.981 &       23.403  &       0.01315 \\
0889    &       20.281  &       3.826   &       0.01776 &       22889   &       180.352 &       -18.832 &       0.01657 \\
3146    &       23.316  &       30.746  &       0.04004 &       21578   &       184.771 &       5.819   &       0.02551 \\
20173   &       32.009  &       35.463  &       0.02230 &       23494   &       186.144 &       7.186   &       0.01566 \\
20828   &       32.556  &       -0.198  &       0.03283 &       3390    &       186.307 &       12.662  &       0.12703 \\
1850    &       34.570  &       -73.938 &       0.03450 &       21781   &       188.338 &       70.765  &       0.04260 \\
21766   &       34.963  &       -6.148  &       0.01781 &       0497    &       188.965 &       12.498  &       0.01774 \\
3378    &       37.592  &       -60.554 &       0.15114 &       2105    &       194.340 &       26.898  &       0.01816 \\
3020    &       40.277  &       -8.315  &       0.01193 &       2103    &       194.450 &       27.402  &       0.04238 \\
3334    &       40.876  &       32.421  &       0.30850 &       0560    &       195.646 &       -2.307  &       0.02117 \\
3093    &       50.413  &       -37.128 &       0.03664 &       21170   &       197.706 &       57.658  &       0.05172 \\
2426    &       50.602  &       -37.160 &       0.08201 &       23576   &       198.539 &       -16.381 &       0.03641 \\
20052   &       54.544  &       0.318   &       0.03535 &       0447    &       201.280 &       -38.507 &       0.34308 \\
3398    &       55.118  &       -18.574 &       0.02939 &       1359    &       202.274 &       58.447  &       0.02694 \\
1709    &       56.078  &       24.589  &       0.01932 &       20318   &       203.157 &       -31.798 &       0.07504 \\
22693   &       58.565  &       -59.036 &       0.01911 &       20177   &       204.294 &       51.931  &       0.03124 \\
20953   &       61.904  &       -12.364 &       0.03154 &       0406    &       209.029 &       18.395  &       0.06527 \\
2502    &       62.600  &       -75.231 &       0.02606 &       0071    &       211.053 &       -33.858 &       0.05082 \\
1981    &       63.084  &       -28.533 &       0.01731 &       3368    &       211.633 &       25.134  &       0.00993 \\
2343    &       68.093  &       -13.263 &       0.03117 &       3369    &       211.732 &       25.011  &       0.01606 \\
1292    &       72.174  &       -66.052 &       0.04434 &       3113    &       213.205 &       -34.298 &       0.02489 \\
3116    &       73.702  &       -10.254 &       0.02499 &       1766    &       213.722 &       36.205  &       0.08474 \\
2288    &       83.290  &       -62.426 &       0.01029 &       20700   &       222.450 &       8.906   &       0.03159 \\
23586   &       86.227  &       -25.738 &       0.01752 &       2372    &       227.778 &       70.718  &       0.04958 \\
21432   &       117.482 &       55.862  &       0.03641 &       0907    &       233.187 &       32.712  &       0.01681 \\
1930    &       122.110 &       -76.477 &       0.01407 &       22703   &       251.264 &       57.630  &       0.03171 \\
3305    &       122.209 &       20.932  &       0.02117 &       3193    &       251.927 &       34.955  &       0.06984 \\
3457    &       139.828 &       -11.985 &       0.01521 &       24820   &       260.420 &       57.876  &       0.02510 \\
1618    &       144.532 &       71.116  &       0.01495 &       3165    &       262.827 &       6.031   &       0.01854 \\
3085    &       145.470 &       46.854  &       0.02163 &       0204    &       314.015 &       -4.524  &       0.01769 \\
1679    &       145.594 &       46.982  &       0.01284 &       1823    &       322.931 &       -42.876 &       0.02355 \\
1777    &       146.697 &       9.805   &       0.26240 &       24447   &       324.265 &       -63.133 &       0.03605 \\
21197   &       149.694 &       2.264   &       0.03038 &       21444   &       330.691 &       18.841  &       0.02231 \\
2316    &       152.129 &       12.202  &       0.01674 &       2844    &       333.578 &       -10.265 &       0.03685 \\
0434    &       156.059 &       4.192   &       0.00467 &       21706   &       334.159 &       -36.817 &       0.03245 \\
24309   &       161.450 &       4.364   &       0.08199 &       22786   &       339.026 &       34.182  &       0.02820 \\
23209   &       162.955 &       57.546  &       0.01016 &       2052    &       344.267 &       -43.337 &       0.03379 \\
22477   &       163.180 &       10.547  &       0.01799 &       2159    &       349.022 &       -2.428  &       0.04161 \\
0813    &       163.192 &       57.355  &       0.01775 &       1911    &       349.555 &       -42.193 &       0.05136 \\
22475   &       163.281 &       10.707  &       0.03708 &       21587   &       352.266 &       14.865  &       0.04891 \\
3060    &       164.630 &       1.634   &       0.02172 &               &               &               &               \\

\hline
\end{tabular}
    \label{table:prov}
\end{table}
\twocolumn

\end{appendix}

\end{document}